\documentclass[journal]{IEEEtran}
\usepackage{enumerate}
\usepackage{amsmath}
\usepackage{amscd}
\usepackage{dsfont}
\usepackage{mathtools}
\usepackage{amssymb}
\usepackage{latexsym}
\usepackage{array}
\usepackage{tabularx}
\usepackage{setspace}
\usepackage{tikz}
\usetikzlibrary{calc}
\usepackage{graphicx,subfigure}
\usepackage{bbm}
\usepackage[numbers, square, comma, compress]{natbib}
\newtheorem{lem}{Lemma}
\newtheorem{corol}{Corollary}
\newtheorem{theorem}{Theorem}
\newtheorem{prop}{Proposition}

\newtheorem{deft}{Definition}
\newtheorem{rem}{Remark}
\providecommand{\mb}[1]{\mathbf{#1}}

\newcommand{\I}{\mathbb{I}}
\newcommand{\Z}{\mathsf{Z}}
\newcommand{\M}{\mathsf{M}}
\newcommand{\V}{\mathbb{V}}

\newcommand{\R}{\mathbb{R}}

\newcommand{\X}{\mathsf{X}}
\newcommand{\Y}{\mathsf{Y}}
\newcommand{\W}{\mathsf{W}}
\newcommand{\K}{\mathsf{K}}
\newcommand{\U}{\mathsf{U}}
\newcommand{\D}{\mathbb{D}}

\renewcommand{\H}{\mathbb{H}}
\renewcommand{\S}{\mathsf{S}}

\newcommand {\aplt} {\ {\raise-.5ex\hbox{$\buildrel<\over{\mbox{\scriptsize $\sim$}}$}}\ }
\providecommand{\abs}[1]{\ensuremath{\left\lvert #1 \right\rvert}}
\providecommand{\norm}[1]{\ensuremath{\left\Vert #1 \right\Vert}}
\providecommand{\vv}[1]{\textquotedblleft #1\textquotedblright}
\providecommand{\nint}[1]{\ensuremath{\left\lfloor #1 \right \rceil}}
\providecommand{\floor}[1]{\ensuremath{\left\lfloor #1 \right \rfloor}}
\providecommand{\ceil}[1]{\ensuremath{\left\lceil #1 \right \rceil}}

\renewcommand{\IEEEQED}{\IEEEQEDopen}
\DeclareMathOperator*{\av}{av}

\DeclareMathOperator*{\Mod}{\,mod}

\newcommand{\KL}{KL}

\makeatletter \g@addto@macro\@floatboxreset\centering \makeatother

\definecolor{myred}{rgb}{0.8,0,0}
\definecolor{mygreen}{RGB}{34,139,34}
\definecolor{myblue}{RGB}{0,0,205}

\providecommand{\myalert}[1]{\textcolor[RGB]{0,0,205}{#1}}
\providecommand{\redalert}[1]{\textcolor[rgb]{0.8,0,0}{#1}}
\providecommand{\greenalert}[1]{\textcolor[rgb]{0.0,0.5,0.0}{#1}}

\begin{document}


\title{Optimal rate-limited secret key generation from Gaussian sources using lattices}

\author{Laura Luzzi, Cong Ling and Matthieu R. Bloch
	\thanks{
	The work of L. Luzzi was supported in part 
	 by CY Initiative of Excellence ``Investissements d'Avenir'' under grants AAP2017 Lattice Hashing and   ANR-16-IDEX-0008.
	The work of C. Ling was supported in part by the Engineering and Physical
Sciences Research Council (EPSRC) under Grant No. EP/S021043/1.
The work of M. Bloch was supported in part by the National Science Foundation under awards 1955401 and 2148400.
This work was presented in part at the IEEE International Symposium on Information Theory (ISIT 2013), Istanbul, Turkey \cite{LLB13}, and in part at the International Zurich Seminar on Communications (IZS 2018) \cite{IZS2018}.
}

\thanks{L. Luzzi is with ETIS, UMR 8051  (CY Cergy Paris Universit\'{e}, ENSEA, CNRS), 95014 Cergy-Pontoise, France (e-mail: laura.luzzi@ensea.fr). 

C. Ling is with the Department of Electrical and Electronic Engineering,
Imperial College London, London SW7 2AZ, U.K. (e-mail: cling@ieee.org). 

M. R. Bloch is with School of Electrical and Computer Engineering,
Georgia Institute of Technology, Atlanta, Georgia (email: matthieu.bloch@ece.gatech.edu).
}

}

\maketitle

\begin{abstract}
We propose a lattice-based scheme for secret key generation from Gaussian sources in the presence of an eavesdropper, and show that it achieves the strong secret key capacity in the case of degraded source models, as well as the optimal secret key / public communication rate trade-off. 
The key ingredients of our scheme are 
the use of the modulo lattice operation
to extract the channel intrinsic randomness, based on the notion of flatness factor, together with a randomized lattice quantization technique to quantize the continuous source. Compared to previous works, we introduce two new notions of flatness factor based on $L^1$ distance and KL divergence, respectively, which might be of independent interest. We prove the existence of secrecy-good lattices under $L^1$ distance and KL divergence, whose $L^1$ and KL flatness factors vanish for volume-to-noise ratios up to $2\pi e$. This improves upon the volume-to-noise ratio threshold $2\pi$ of the $L^{\infty}$ flatness factor.
\end{abstract}


\section{Introduction}

Secret key generation (also known as key agreement) at the physical layer was first investigated by Maurer \cite{Maurer93a} and Ahlswede and Csisz\'ar \cite{AC_IT93}, who showed that correlated observations of noisy phenomena could be used to distill secret keys by exchanging information over a public channel. In recent years, this subject has received considerable attention in literature (see, e.g., \cite{ChouTzu-Han15,ChanChung18,TyagiHimanshu15,IwamotoMitsugu18,GohariAmin20,LiCheukTing21}). The setup has been extended to the vector case \cite{LiuJingbo16,KhistiAshish16}, the
multi-terminal case \cite{ChanChung19,TyagiHimanshu17,ChanChung14,GohariAmin10}, the quantum case \cite{Seshadreesan16}
and the case with feedback \cite{BassiGerman19}. Second-order asymptotics have been derived in \cite{HayashiMasahito16, Poostindouz21}. Code constructions for the discrete memoryless case have been proposed, e.g. \cite{MuramatsuJun12, chou2015secretkey}. 

Most existing secret key generation schemes rely heavily on the assumption of discrete random sources over finite or countable alphabets. In order to apply these techniques to wireless communications, it is necessary to extend the key generation framework to the case of continuous sources, such as Gaussian sources \cite{LiuJingbo16,WaOh10,KhistiAshish12,Nitinawarat_Narayan12}\footnote{An extension of the key distillation framework to quantum Gaussian states has also been considered \cite{navascues2005,lami2020}.}. 
In \cite{Nitinawarat_Narayan12}, the authors study a multi-terminal scenario for secret key generation in the special case for which the eavesdropper only has access to the public channel. Beside providing a characterization of the optimal strongly secret key rate, the authors show that this optimal rate can be achieved using lattice codes (for information reconciliation only).

We consider here the problem of secret key generation between two terminals, Alice and Bob, who observe correlated Gaussian sequences $\X^n$ and $\Y^n$, in the presence of an eavesdropper, Eve, who also obtains a correlated sequence $\Z^n$. For simplicity, we suppose that a single round of unidirectional public communication takes place in order to establish the key.
Our main contribution is to show that, in the case of a degraded source model, the strong secret key capacity can be achieved by a complete lattice-coding scheme considerably different from and perhaps simpler than \cite{Nitinawarat_Narayan12}\footnote{The scheme in \cite[Section IV-B]{Nitinawarat_Narayan12} requires the repetition of a dithered quantization and public communication step over $N$ blocks, each of dimension $n$. This is needed to achieve strong secrecy from weak secrecy by using the technique in \cite{Maurer_Wolf}. In contrast, our scheme achieves strong secrecy with a single block and bounds the mutual information using the variational distance, as in \cite{Csiszar96}.}. 
This extends our previous work \cite{LLB13}, in which it was shown that a secret key rate up to half a nat from the optimal was achievable.

Typically, secret key generation consists of two distinct procedures: \emph{information reconciliation}, in which public messages are exchanged to ensure that Alice and Bob can construct the same data sequence with vanishing error probability, and \emph{privacy amplification} to extract from this shared sequence a secret key that is statistically independent from Eve's observation and from the public messages.
\subsubsection*{Privacy amplification and randomness extraction} Our privacy amplification strategy is based on the concept of \emph{channel intrinsic randomness},
or the maximum bit rate that can be extracted from a channel output independently of its input  \cite{Bloch_Intrinsic_Randomness, MuKoMu03, Hayashi11}.
One can show that the reduction modulo a suitable lattice can be used
to extract the intrinsic randomness\footnote{See the discussion in the preprint version of this paper \cite{secret_key_preprint}.}.
Although our main objective in this paper is to solve the problem of privacy amplification, this technique 
is 
an intriguing result in its own right, which could have other applications.
\subsubsection*{The flatness factor and its variants} In our previous work \cite{LLB13}, we provided a characterization of the class of lattices that are good for randomness extraction, which was based on a computable parameter, the \emph{flatness factor}, measuring the $L^{\infty}$ distance between the \vv{folded} Gaussian distribution modulo the lattice and the uniform distribution on the corresponding fundamental region.
The concept of flatness factor is related to the smoothing parameter used in lattice-based cryptography \cite{Micciancio_Regev}, and was first introduced in \cite{BelfioreITW11} in the context of physical-layer network coding. In \cite{LLBS}, two of the authors also showed the relevance of the flatness factor for secrecy and introduced the notion of \emph{secrecy-good lattices} for the wiretap channel. 
In this work, we consider two extended notions of flatness factor by which the $L^{\infty}$ distance is replaced respectively by the $L^1$ distance and the Kullback-Leibler (KL) divergence. These new flatness conditions are satisfied by a wider range of variance parameters, resulting in improved volume conditions for the chain of lattices under consideration, which allows us to achieve the secret key capacity. 
The existence of lattices with vanishing $L^1$ and KL flatness factors follows by leveraging an existence result for resolvability codes for regular channels \cite{Hayashi_Matsumoto2016}.
We note that the $L^1$ smoothing parameter was already considered in \cite{Chung2013,Dadush_Regev_2016}, while $L^1$ and KL flatness factors were used implicitly earlier in  \cite[p. 1656]{LiuLing18}. An upper bound on the $L^1$ flatness factor based on the Cauchy-Schwarz inequality was given in \cite{Mirghasemi14}. The independent work \cite{Thomas22} studied $L^1$ smoothing parameters both for lattices and for codes, also based on the Cauchy-Schwarz inequality. Our approach bypasses the Cauchy-Schwarz inequality, therefore leading to a tighter bound than \cite{Mirghasemi14}. We note however that \cite{Thomas22} obtained a bound on the $L^1$ smoothing parameter as tight as that in this paper, by decomposing the discrete Gaussian distribution into a convex combination of uniform ball distributions. The smoothing parameter is of fundamental importance in lattice and code-based cryptography \cite{Thomas22}, so our method for the $L^1$ flatness factor may also be useful in these areas.

\subsubsection*{Information reconciliation and Wyner-Ziv coding} Our strategy for information reconciliation follows the outline of \cite{WaOh10,Nitinawarat_Narayan12}: first, the source $\X^n$ is vector quantized; then, a public message is generated in the manner of Wyner-Ziv coding, so that Bob can decode the quantized variable using the sequence $\Y^n$ as side information.  The existence of good nested lattices for Wyner-Ziv coding has been established in \cite{Zamir02} (see also \cite{LCLX06,LingCOM12}). We show that this construction is compatible with the secrecy-goodness property to conclude our existence proof.
\subsubsection*{Randomized quantization technique} Unlike our previous work \cite{LLB13}, the quantization performed at Alice's side is not deterministic. We introduce a new \emph{randomized quantization} step inspired by the randomized rounding technique in \cite{Peikert}. Essentially, this technique allows to round a continuous Gaussian into a \emph{discrete Gaussian distribution} with slightly larger variance, provided that the $L^{\infty}$ flatness factor of the lattice is small. We partially extend the result of \cite{Peikert} under an $L^1$ flatness factor criterion. We show that randomized quantization with uniform dithering (where the dither is known by all parties, including the eavesdropper) achieves the optimal trade-off between public communication rate and secret key rate established in \cite{WaOh10}. The dithering technique has been used to achieve capacity in literature \cite{ErezZamir04,Antonio}. Besides, the discrete Gaussian distribution is widely used in lattice coding \cite{LLBS} and lattice-based cryptography \cite{Chung2013,Peikert}. However, its application to quantization is new, to the best of our knowledge.

\subsubsection*{Relation to fuzzy extractors} 
Fuzzy extractors \cite{Dodis} allow to extract a secret key from a noisy measurement, which means that it is resilient to small measurement errors. Fuzzy extractors for continuous signals were proposed in \cite{Verbitskiy,Linnartz}. Our proposed lattice code is also robust to measurement errors, thanks to its channel coding component of Wyner-Ziv coding. A notable difference is that min-entropy is used to measure the available randomness in fuzzy extractors, while Shannon entropy is used in our key generation model. 
Moreover, for fuzzy extractors the measurement error is assumed to have bounded Hamming weight or Euclidean norm, while in our model it follows a Gaussian distribution.

\subsubsection*{Organization}
This paper is organized as follows. In Section \ref{section_preliminaries} we provide basic definitions about lattices and recall the notion  of $L^{\infty}$ flatness factor. In Section \ref{flatness_factor_section} we define a new $L^1$ variant of the flatness factor, which allows us to define the notion of $L^1$ secrecy-good lattices. 
In Section \ref{SKG_section}, we introduce the Gaussian source model, describe our lattice-based secret key generation scheme and prove our main result.
Finally, in Section \ref{conclusion} we offer some conclusions and perspectives.
 For ease of reading, the additional technical tools needed to prove the existence of good nested lattices are presented in the Appendix.
 More precisely, Appendix 
\ref{appendix_resolvability} summarizes some relevant results on the existence of resolvability codes for regular channels. Appendix \ref{Appendix_KL} presents the KL flatness factor and its properties.
The existence of lattices that are KL secrecy-good and, consequently, also $L^1$ secrecy-good is proven in Appendix \ref{proof_VNR_L1}. Finally, the existence of the sequences of nested lattices required in our key generation scheme is proven in Appendix \ref{existence_section}.  

\section{Preliminaries on lattices and the $L^{\infty}$ flatness factor} \label{section_preliminaries}
\paragraph*{Notation}
All logarithms in this paper are assumed to be natural logarithms, and information is measured in nats. Given a set $A$, the notation $\mathcal{U}_{A}$ stands for the uniform distribution over $A$. The notation $\mathbb{F}_p$ refers to the finite field of order $p$.
We denote the variational distance between two (discrete or continuous) distributions $p,q$ by $\V(p,q)$, and their KL divergence by $\mathbb{D}(p\|q)$.

In this section, we recall some well-known properties of lattices as well as the notion of flatness factor based on $L^{\infty}$ distance.


An $n$-dimensional {lattice} $\Lambda$ in the Euclidean space
$\mathbb{R}^{n}$ is the discrete set defined by
\begin{equation*}
\Lambda=\mathcal{L}\left( \mathbf{B}\right) =\left\{ \mathbf{Bx}\text{ : }\mathbf{x\in }\text{ }%
\mathbb{Z}^{n}\right\}
\end{equation*}%
where the columns of the basis matrix $\mathbf{B=}\left[ \mathbf{b}_{1}\cdots \mathbf{b}_{n}\right]
$ are linearly independent.

Given a lattice $\Lambda$, its dual lattice $\Lambda^*$ is defined as the set of vectors $\lambda^*$ in $\R^n$ such that $\langle \lambda^*,\lambda \rangle \in \mathbb{Z}$ for all $\lambda \in \Lambda$.

A measurable set~$\mathcal{R}(\Lambda)\subset \mathbb{R}^n$ is called a fundamental region of the
lattice~$\Lambda$ if the disjoint union~$\cup_{\lambda \in \Lambda} (\mathcal{R}(\Lambda)+\lambda) = \R^n$.
Examples of fundamental regions
include the fundamental parallelepiped $\mathcal{P}(\Lambda)$ and the Voronoi region $\mathcal{V}(\Lambda)$. All the fundamental regions have equal
volume~$V(\Lambda)$.

Given a lattice $\Lambda$ and a fundamental region $\mathcal{R}(\Lambda)$, any point $\mathbf{x} \in \R^n$ can be written uniquely as a sum
$$ \mb{x}=\lambda+\bar{\mb{x}},$$
where $\lambda \in \Lambda$ and $\bar{\mb{x}} \in \mathcal{R}(\Lambda)$. The vector $\lambda$ is the quantization of $\mathbf{x}$ with respect to $\mathcal{R}(\Lambda)$ and is denoted as $Q_{\mathcal{R}(\Lambda)}(\mb{x})$, where boundary points are decided systematically. Thus we define 
\begin{equation} \label{eq:mod}
[\mb{x}] \Mod \mathcal{R}(\Lambda)=\mb{x}-Q_{\mathcal{R}(\Lambda)}(\mb{x})=\bar{\mb{x}}.
\end{equation}
In particular, for any $\mathbf{x} \in \R^n$, the nearest-neighbor quantizer associated with $\Lambda$ is given by $$Q_{\Lambda}(\mathbf{x})=Q_{\mathcal{V}(\Lambda)}(\mb{x})=\arg\min_{\lambda \in\Lambda}\|\lambda-\mathbf{x}\|$$
where ties are broken systematically.
Note that $\mathbf{x} \Mod \mathcal{V}(\Lambda)=\mathbf{x}-Q_{\Lambda}(\mathbf{x})$.
The modulo lattice operation satisfies the distributive law \cite[Proposition 2.3.1]{Zamir_book}, i.e., $\forall \lambda \in \Lambda$
 \begin{equation} \label{distributive_law}
 [\mb{x} + \lambda] \Mod \mathcal{R}(\Lambda)=[\mb{x}] \Mod \mathcal{R}(\Lambda).
 \end{equation}
 
The following property \cite[equation (35)]{Nazer11} will also be used in the paper: given two lattices $\Lambda \subseteq \Lambda_1$, $\mb{x} \in \R^n$, and a fundamental region $\mathcal{R}(\Lambda)$,
\begin{equation} \label{NG_35}
[Q_{\Lambda_1}(\mb{x})] \Mod \mathcal{R}(\Lambda)= [Q_{\Lambda_1}([\mb{x}] \Mod \mathcal{R}(\Lambda))] \Mod \mathcal{R}(\Lambda).
\end{equation}

Given a sublattice $\Lambda' \subset \Lambda$, the quotient group $\Lambda / \Lambda'$ is defined as the group of distinct cosets $\lambda+\Lambda'$ for $\lambda \in \Lambda$. It can be identified by a set of coset representatives $\Lambda \cap \mathcal{R}(\Lambda')$, where $\mathcal{R}(\Lambda')$ is any fundamental region of $\Lambda'$. Furthermore,  $\mathcal{R}(\Lambda')$ can be written as a disjoint union of translates of any fundamental region $\mathcal{R}(\Lambda)$ as follows \cite[equation (8.33)]{Zamir_book}:
\begin{equation} \label{Zamir_8_33}
\mathcal{R}(\Lambda')=\bigcup_{\lambda \in \Lambda \cap \mathcal{R}(\Lambda')}\left(\left[\lambda+\mathcal{R}(\Lambda)\right]\Mod \mathcal{R}(\Lambda')\right).
\end{equation}

Suppose that $\X^n$ is an $n$-dimensional i.i.d. Gaussian random variable of variance $\sigma^2$ with distribution
\begin{equation*}
 f_{\sigma}(\mathbf{x})=\frac{1}{(\sqrt{2\pi}\sigma)^n}e^{- \frac{\|\mathbf{x}\|^2}{2\sigma^2}},
\end{equation*}
for $\mathbf{x} \in\R^n$. 
The following useful property characterizing the product of Gaussian distributions was proven in \cite[Fact 2.1]{Peikert}\footnote{Note that although the statement in \cite{Peikert} refers to (unnormalized) Gaussian functions, one can check that it also holds for Gaussian distributions.}:
\begin{lem} \label{Fact1}
Given $\sigma_1, \sigma_2>0$, let $\sigma$ and $\bar{\sigma}$ be such that $\sigma^2=\sigma_1^2+\sigma_2^2$, and $\frac{1}{\bar{\sigma}^2}=\frac{1}{\sigma_1^2}+\frac{1}{\sigma_2^2}$. Moreover, let $\mb{c}_1,\mb{c}_2 \in \R^n$, and $\bar{\mb{c}}=\frac{\bar{\sigma}^2}{\sigma_1^2}\mb{c}_1+\frac{\bar{\sigma}^2}{\sigma_2^2}\mb{c}_2$.  Then $\forall \mb{x} \in \R^n$,
$$f_{\sigma_1}(\mb{x}-\mb{c}_1)f_{\sigma_2}(\mb{x}-\mb{c}_2)=f_{\sigma}(\mb{c}_1-\mb{c}_2)f_{\bar{\sigma}}(\mb{x}-\bar{\mb{c}}).$$
\end{lem}

Given a lattice $\Lambda$, we define the $\Lambda$-periodic function
\begin{equation}\label{Guass-function-lattice}
  f_{\sigma,\Lambda}(\mathbf{x})=\frac{1}{(\sqrt{2\pi}\sigma)^n}
\sum_{\lambda \in \Lambda} e^{-
    \frac{\|\mathbf{x}+\lambda\|^2}{2\sigma^2}},
\end{equation}
for all $\mathbf{x} \in\R^n$. We denote by    $f_{\sigma,\mathcal{R}(\Lambda)}={f_{\sigma,\Lambda}}_{|\mathcal{R}(\Lambda)}$ its restriction to the fundamental region $\mathcal{R}(\Lambda)$. Note that $f_{\sigma,\mathcal{R}(\Lambda)}$ is the probability density of $\bar{\X}^n=[\X^n] \Mod \mathcal{R}(\Lambda)$. Given $\mb{c} \in \R^n$, we will also use the notation
$$f_{\sigma,\Lambda,\mb{c}}(\mb{x})=f_{\sigma,\Lambda}(\mb{x}-\mb{c})$$
to denote a shifted $\Lambda$-periodic function.

Given an $n$-dimensional lattice $\Lambda$ in $\R^n$ and a vector $\mathbf{c} \in \R^n$, we define the \emph{discrete Gaussian distribution} over $\Lambda$
centered at $\mathbf{c}$ as the following discrete
distribution taking values in $\lambda \in \Lambda$:
\[
D_{\Lambda,\sigma,\mathbf{c}}(\lambda)=\frac{f_{\sigma,\mathbf{c}}(\mathbf{\lambda})}{f_{\sigma,\Lambda}(\mb{c})} \quad \forall \lambda \in \Lambda.
\]
We write $D_{\Lambda,\sigma}=D_{\Lambda,\sigma,\mathbf{0}}$. 
Following Peikert \cite[Section 4.1]{Peikert}, we introduce the notion of randomized rounding with respect to $\Lambda$: 
\begin{deft}[Randomized rounding]
Given an input vector $\mb{x} \in \R^n$, we define the random variable 
\begin{equation} \label{randomized_rounding_definition}
\nint{\mb{x}}_{\Lambda, \sigma} \sim D_{\Lambda, \sigma,\mb{x}}.
\end{equation}
\end{deft}
Note that $\nint{\mb{x}}_{\Lambda, \sigma}$ is a discrete random variable taking values in $\Lambda$. 

In essence, randomized rounding consists in sampling from a lattice Gaussian distribution centered at $\mathbf{x}$. There exist several algorithms for this task. In particular, it was proven in~\cite{GPV} that Klein's
algorithm~\cite{Klein} samples from a distribution very close to~$D_{\Lambda,\sigma,{\bf x}}$
when~$\sigma$ is sufficiently large. A new algorithm was given in~\cite{ZhengWangTIT15} which overcomes the restriction on~$\sigma$.

\begin{deft} [$L^{\infty}$  Flatness factor \cite{LLBS}]
For a lattice~$\Lambda$ and for a parameter~$\sigma$, the $L^{\infty}$  flatness factor
is defined by:
\begin{equation*}
\epsilon_{\Lambda}(\sigma)  \triangleq \max_{\mathbf{x} \in
\mathcal{R}(\Lambda)}\abs{
V(\Lambda)f_{\sigma,\Lambda}(\mathbf{x})-1}.
\end{equation*}
\end{deft}

In other words, $\epsilon_{\Lambda}(\sigma)$ characterizes the $L^{\infty}$ distance of $f_{\sigma,\Lambda}(\mathbf{x})$ to the uniform distribution $\mathcal{U}_{\mathcal{R}(\Lambda)}$  over~$\mathcal{R}(\Lambda)$.


%

The $L^{\infty}$  flatness factor is independent of the choice of the fundamental region $\mathcal{R}(\Lambda)$ and can be computed from the theta series of the lattice
\begin{equation} \label{theta_series}
\Theta_{\Lambda}(\tau)=\sum_{\lambda \in \Lambda} e^{-\pi \tau \norm{\lambda}^2}
\end{equation}
using the identity \cite[Proposition 2]{LLBS}
\begin{equation} \label{flatness_factor_expression}
\epsilon_{\Lambda}(\sigma) =  \left(\frac{\gamma_{\Lambda}(\sigma)}{{2\pi}}\right)^{\frac{n}{2}}{
\Theta_{\Lambda}\left({\frac{1}{2\pi\sigma^2}}\right)}-1,
\end{equation}
where $\gamma_{\Lambda}(\sigma) = \frac{
V(\Lambda)^{\frac{2}{n}}}{\sigma^2}$ is the volume-to-noise ratio (VNR). Moreover, the following relation holds between the flatness factor of $\Lambda$ and the theta series of its dual lattice $\Lambda^*$ \cite[Corollary 1]{LLBS}:
\begin{equation} \label{theta_dual}
\Theta_{\Lambda^*}(2\pi\sigma^2)=\epsilon_{\Lambda}(\sigma)+1.
\end{equation}

\begin{rem} \label{monotonicity}
We have shown in \cite{LLBS} that $\epsilon_{\Lambda}$ is a monotonically decreasing function, i.e., for $\sigma < \sigma'$, we have $\epsilon_{\Lambda}(\sigma') \leq  \epsilon_{\Lambda}(\sigma)$.
\end{rem}

The notion of secrecy-goodness
characterizes lattice sequences whose $L^{\infty}$ flatness factors vanish exponentially fast
as $n \to \infty$.

\begin{deft}[Secrecy-good lattices under $L^{\infty}$  flatness factor \cite{LLBS}] \label{secrecy_goodness}
A sequence of lattices $\Lambda^{(n)}$
is \emph{secrecy-good} under the $L^{\infty}$  flatness factor if
$\epsilon_{\Lambda^{(n)}}(\sigma)  = e^{-\Omega(n)}$
for all fixed $\gamma_{\Lambda^{(n)}}(\sigma)<2\pi$.
\end{deft}

In \cite{LLBS} we have proven the existence of sequences of secrecy-good lattices under $L^{\infty}$  flatness factor  as long as
\begin{equation} \label{secrecy_condition}
\gamma_{\Lambda}(\sigma)<2\pi.
\end{equation}


\section{Secrecy-good lattices under an $L^1$ flatness factor condition} \label{flatness_factor_section}
In this section, we introduce a weaker notion of flatness based on the $L^1$ distance and study its properties.
\begin{deft}
Given a lattice $\Lambda$, a fundamental region $\mathcal{R}(\Lambda)$ and $\sigma>0$, we define the \emph{$L^1$ flatness factor} as follows:
\begin{equation} \label{L1_flatness_factor}
\epsilon^1_{\Lambda}(\sigma)=\int_{\mathcal{R}(\Lambda)}\abs{f_{\sigma,\Lambda}(\mb{x})-\frac{1}{V(\Lambda)}} d\mb{x}= \V(f_{\sigma,\mathcal{R}(\Lambda)},\mathcal{U}_{\mathcal{R}(\Lambda)}).
\end{equation}
\end{deft}
Similarly to the $L^{\infty}$ flatness factor,  the $L^1$ flatness factor does not depend on the choice of the fundamental region. Moreover, it is shift-invariant, i.e. $\forall \mb{c} \in \R^n$,
\begin{equation} \label{L1_shift_invariant}
\epsilon^1_{\Lambda}(\sigma)=\mathbb{V}({f_{\sigma,\Lambda,\mb{c}}}_{|\mathcal{R}(\Lambda)},\mathcal{U}_{\mathcal{R}(\Lambda)}).
\end{equation}%
\begin{rem} 
For any lattice $\Lambda$, $\forall \sigma>0$, we have
$\epsilon^1_{\Lambda}(\sigma) \leq \epsilon_{\Lambda}(\sigma).$
\end{rem}

The $L^1$ flatness factor is related to the $L^1$ smoothing parameter, which was discussed in \cite{Chung2013,Dadush_Regev_2016}. 

The following Lemma confirms the intuition that folded additive Gaussian noise with larger variance looks more uniform:

\begin{lem} \label{L1_monotonic}
	The $L^1$ flatness factor is monotonic, i.e. for any lattice $\Lambda$, $\forall \sigma'>\sigma$,
	$$\epsilon^{1}_{\Lambda}(\sigma') \leq \epsilon^{1}_{\Lambda}(\sigma).$$
	\end{lem}
\begin{IEEEproof}
	Suppose that $\W^n \sim \mathcal{N}(0,\sigma^2I_n)$, and let $\X^n= \W^n \Mod \mathcal{R}(\Lambda) \sim f_{\sigma,\mathcal{R}(\Lambda)}$. Given $\sigma_0>0$, let $\W_0^n \sim \mathcal{N}(0,\sigma_0^2I_n)$ and consider
\begin{align*}
&\Y^n=[\X^n+\W_0^n] \Mod \mathcal{R}(\Lambda)\\
&=[[\W^n] \Mod \mathcal{R}(\Lambda))+\W_0^n] \Mod \mathcal{R}(\Lambda)\\
&\stackrel{(a)}{=}[\W^n + \W_0^n] \Mod \mathcal{R}(\Lambda) \sim f_{\sqrt{\sigma^2+\sigma_0^2},{\mathcal{R}(\Lambda)}},
\end{align*}
where (a) follows from the distributive property (\ref{distributive_law}). Now consider the random variable $\mathsf{U}^n \sim \mathcal{U}_{\mathcal{R}(\Lambda)}$. By the Crypto Lemma \cite[Lemma 4.1.1]{Zamir_book}, 
$$[\U^n+\W_0^n] \Mod \mathcal{R}(\Lambda) \sim \mathcal{U}_{\mathcal{R}(\Lambda)}.$$
Then using the data processing inequality for the variational distance \cite[Lemma 8]{Bloch_Laneman},
\begin{align*}
&\epsilon^{1}_{\Lambda}\left(\sqrt{\sigma^2+\sigma_0^2}\right)=
\mathbb{V}\left(f_{\sqrt{\sigma^2+\sigma_0^2},\mathcal{R}(\Lambda)},\mathcal{U}_{\mathcal{R}(\Lambda)}\right)=
\V(\Y^n,\mathsf{U}^n) \\
&\leq \V(\X^n, \mathsf{U}^n)= 
 \mathbb{V}(f_{\sigma,\mathcal{R}(\Lambda)},\mathcal{U}_{\mathcal{R}(\Lambda)})=\epsilon^{1}_{\Lambda}(\sigma).
 \end{align*}
 Since this is true for any $\sigma_0>0$, the conclusion follows.
%
	\end{IEEEproof}
\begin{rem} \label{remark_nested_lattices}
For any pair of nested lattices $\Lambda'\subset \Lambda$, $\forall \sigma>0$, we have $\epsilon_{\Lambda}^1(\sigma)\leq \epsilon^1_{\Lambda'}(\sigma)$.
\end{rem}
\begin{IEEEproof}
Given fundamental regions $\mathcal{R}(\Lambda)$, $\mathcal{R}(\Lambda')$, the statement follows easily by noting that
{\allowdisplaybreaks
\begin{align}
&\epsilon^1_{\Lambda}(\sigma)=\int_{\mathcal{R}(\Lambda)} \bigg|\frac{1}{V(\Lambda)} -\sum_{\smash{\tilde{\lambda} \in \Lambda/\Lambda'}}f_{\sigma,\Lambda'}(\mb{u}+\tilde{\lambda})\bigg|d\mb{u} \notag\\
&\leq \sum_{\smash{\tilde{\lambda} \in \Lambda/\Lambda'}} \int_{\mathcal{R}(\Lambda)} \abs{\frac{1}{V(\Lambda')} -f_{\sigma, \Lambda'}(\mb{u}+\tilde{\lambda})}d\mb{u} \notag\\
&=\int_{\mathcal{R}(\Lambda')}\abs{\frac{1}{V(\Lambda')}-f_{\sigma,\Lambda'}(\mb{v})}d\mb{v}=\epsilon^1_{\Lambda'}(\sigma). \tag*{\IEEEQED}
\end{align}
}
\let\IEEEQED\relax%
\end{IEEEproof}

We will next show that lattices that are good for secrecy in the
$L^1$ 
sense exist and that the corresponding volume condition is less stringent than the condition (\ref{secrecy_condition}) for secrecy-goodness based on the $L^{\infty}$ metric. 

\begin{deft} \label{L1_secrecy_good}
	A sequence of lattices $\{\Lambda^{(n)}\}$ is \emph{$L^1$ secrecy-good} if for all fixed $\gamma_{\Lambda^{(n)}}(\sigma) < 2\pi e$, $\forall c>0$, $\epsilon^{1}_{\Lambda^{(n)}}(\sigma) =o\left(\frac{1}{n^c}\right)$, i.e., the $L^1$ flatness factor vanishes super-polynomially.
\end{deft}

 The following theorem, which was presented in \cite{IZS2018}, is the first main result of this paper:

\begin{theorem} \label{VNR_L1}
If $\gamma_{\Lambda}(\sigma)<2 \pi e$ is fixed, then there exists a sequence $\{\Lambda^{(n)}\}$ of lattices which are $L^1$-secrecy good.
\end{theorem}

The proof of Theorem \ref{VNR_L1} is given in Appendix \ref{proof_VNR_L1}. Our proof is information-theoretic and does not require the knowledge of the theta series, in contrast to the $L^{\infty}$ flatness factor. We outline the key ideas here. In order to show the existence of a sequence of lattices $\Lambda^{(n)}$ such that  
$\epsilon^{1}_{\Lambda^{(n)}}(\sigma)=\mathbb{V}(f_{\sigma,\mathcal{R}(\Lambda^{(n)})},\mathcal{U}_{\mathcal{R}(\Lambda^{(n)})})\to 0$, we actually prove a stronger result, namely that $\mathbb{D}(f_{\sigma,\mathcal{R}(\Lambda^{(n)})}||\mathcal{U}_{\mathcal{R}(\Lambda^{(n)})})\to 0$. This requires some additional technical tools that are presented in Appendix \ref{Appendix_KL}. 
We build the required lattices using Construction A, and their existence follows from the existence of linear resolvability codes in \cite{Hayashi_Matsumoto2016} (see Appendix \ref{appendix_resolvability} for more details).

\begin{rem}
It is worth mentioning that as soon as the VNR exceeds $2\pi $, the $L^{\infty}$ flatness factor increases exponentially.
In fact, it is easy to see that the bound  $\gamma_{\Lambda}(\sigma)<2 \pi$ is sharp:  the $L^{\infty}$ flatness factor of a lattice cannot vanish for any $\gamma_{\Lambda}(\sigma)>2\pi$. This is simply because \eqref{flatness_factor_expression} implies that
\begin{eqnarray}
\begin{aligned}
\epsilon_{\Lambda}(\sigma) &>\left(\frac{\gamma_{\Lambda}(\sigma)}{{2\pi}}\right)^{\frac{n}{2}}
-1 \notag
\end{aligned}
\end{eqnarray}
since $\Theta_{\Lambda}(\tau) > 1$ for any $\tau>0$.
Thus, as the VNR approaches $2\pi e$, the $L^{\infty}$ flatness factor $\approx e^{n/2}$, but the $L^{1}$ flatness factor can still be brought under control. This demonstrates the advantage of the $L^{1}$ flatness factor.

Also note that the VNR of an $L^1$-secrecy-good lattice approaches $2\pi e$ from below, while that of an AWGN-good lattice approaches $2\pi e$ from above. Recall that the normalized second moment of a quantization-good lattice approaches $1/(2\pi e)$ \cite{Zamir_book}, so all three types of lattices finally share the same VNR threshold $2\pi e$.
\end{rem}

In the following, we discuss the implication of Theorem \ref{VNR_L1} on the smoothing parameter\footnote{We remark that
this definition differs slightly from the one in~\cite{Micciancio_Regev}, where~$\sigma$ is scaled by a constant factor $\sqrt{2\pi}$ (i.e., $s=\sqrt{2\pi}\sigma$).} that is commonly used in lattice-based
cryptography.

\begin{deft} [Smoothing parameter]
For a lattice $\Lambda$ and for $\varepsilon > 0$, the $L^{\infty}$ and $L^{1}$ smoothing parameters
$\eta_{\varepsilon}(\Lambda)$ and $\eta^1_{\varepsilon}(\Lambda)$, respectively, are the smallest $\sigma>0$ such that
$\epsilon_{\Lambda}(\sigma),\epsilon^1_{\Lambda}(\sigma)\leq \varepsilon$.
\end{deft}

Theorem \ref{VNR_L1} implies the existence of lattices whose smoothing parameters $\eta^1_{\varepsilon_n}(\Lambda) \approx \frac{V(\Lambda)^{1/n}}{\sqrt{2\pi e}}$ for a suitable sequence $\varepsilon_n\to 0$. This improves upon the result $\eta_{\varepsilon_n}(\Lambda) \approx \frac{V(\Lambda)^{1/n}}{\sqrt{2\pi }}$.
Using the Cauchy-Schwarz inequality, the following bound was proven in \cite{Mirghasemi14}\footnote{A similar bound was given in \cite{Thomas22} using the statistical distance, which differs from the $L^{1}$ distance by a factor $\frac{1}{2}$.}
\begin{equation}
    \epsilon^1_{\Lambda}(\sigma) \leq \sqrt{\epsilon_{\Lambda}\left(\sqrt{2}\sigma\right)}
\end{equation}
which implies the bound $\eta^1_{\varepsilon}(\Lambda) \leq \frac{V(\Lambda)^{1/n}}{2\sqrt{\pi}}$. However, this bound is not optimal.

\section{Secret key generation} \label{SKG_section}

In this section, we present our system model for secret key generation from correlated Gaussian sources with one-way rate limited communication, in the presence of an eavesdropper, and our proposed key generation protocol based on nested lattices. 

\subsection{System model} \label{section_system_model}
\begin{figure}[bt]
\begin{center}
\begin{footnotesize}
\begin{tikzpicture}[
nodetype1/.style={
	rectangle,
	rounded corners,
	minimum width=10mm,
	minimum height=7mm,
	dashed,
	draw=black,
	text centered
},
nodetype2/.style={
	rectangle,
	rounded corners,
	minimum width=16mm,
	minimum height=7mm,
	text width=16mm,
	text centered,
	draw=black
},
tip2/.style={-latex,shorten >=0.4mm}
]
\matrix[row sep=0.6cm, column sep=0.7cm, ampersand replacement=\&]{
\node (Alice) {\textsc{Alice}}; \& \& \& \node (Bob) {\textsc{Bob}};\\
\node (key_gen1)  [nodetype2] {\textsc{key \\ generation}};  \& \node (quantizer) [nodetype2] {\textsc{quantizer}}; \& \node (source) [draw, nodetype1] {$p_{\X\Y\Z}$}; \& \node (key_gen2) [nodetype2] {\textsc{decoder}}; \\
\& \& \node (Eve) {\textsc{Eve}}; \& \\
};
\draw[->] (key_gen1) edge[tip2] node [right] {$\K$} (Alice);
\draw[->] (key_gen2) edge[tip2] node [right] {$\hat{\K}$} (Bob);
\draw[->] (source) edge[tip2]  node [above] (X) {$\X^n$} (quantizer);
\draw[->] (quantizer) edge[tip2]  node [above=-0.1cm] (XQ) {$\X^n_{Q}$} (key_gen1);
\draw[->] (source) edge[tip2]  node [above] (Y) {$\Y^n$} (key_gen2);
\draw[->] (source) edge[tip2]  node [right] (Z) {$\Z^n$} (Eve);
\draw (Eve.south) node[below=0.4cm] (dot) {};
\draw[-,>=latex] (dot.center) -| node [anchor=south,right,pos=0.9] {$\mathsf{S}$} (key_gen1);
\draw[->,>=latex] (dot.center) -| node [anchor=south, right,pos=0.9] {$\mathsf{S}$} (key_gen2);
\draw[>=latex] (dot.center) edge[tip2]  node [anchor=south, right] {$\mathsf{S}$} (Eve);
\draw (dot.south) node[below] {public channel (noiseless)};
\end{tikzpicture}
\end{footnotesize}
\caption{Secret key generation in the presence of an eavesdropper with communication over a public channel.}
\label{Figure_SecretKey}
\end{center}
\end{figure}
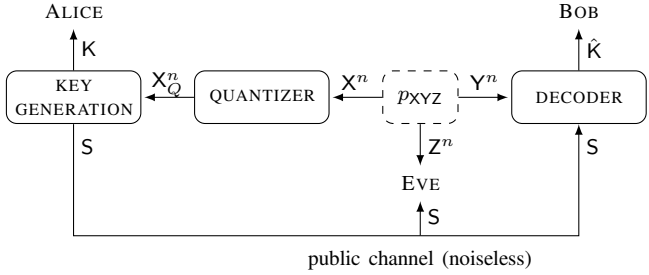

We consider the same model as in \cite{LLB13}, illustrated in Fig. \ref{Figure_SecretKey}, in which Alice, Bob and Eve observe the random variables $\X^n$, $\Y^n$, $\Z^n$ respectively,
generated by an i.i.d. memoryless Gaussian source
$p_{\X\Y\Z}$
whose components are jointly Gaussian with zero mean. The distribution is fully described by the variances $\sigma_x^2$, $\sigma_y^2$, $\sigma_z^2$ and the correlation coefficients $\rho_{xy}$, $\rho_{xz}$, $\rho_{yz}$. We can write~\cite[Eq.~(6)]{WaOh10}:
\begin{equation}\label{XYZ-model}
\left\{
\begin{split}
\X^n &= \rho_{xy}\frac{\sigma_x}{\sigma_y} \Y^n + \mathsf{W}_1^n, \\
\X^n &= \rho_{xz}\frac{\sigma_x}{\sigma_z} \Z^n + \mathsf{W}_2^n,
\end{split}
\right.
\end{equation}
where $\mathsf{W}_1^n$
and $\mathsf{W}_2^n$
are i.i.d. zero-mean Gaussian noise vectors of variances
\begin{equation}
\sigma_1^2=\sigma_x^2(1-\rho_{xy}^2),\quad \sigma_2^2=\sigma_x^2(1-\rho_{xz}^2),
\label{variances}
\end{equation}
respectively, such that $\sigma_2>\sigma_1$. Further,
$\mathsf{W}_1^n$ is independent of
$\Y^n$, and $\mathsf{W}_2^n$
is independent of
$\Z^n$.

We assume that only one round of one-way public communication takes place from Alice to Bob. More precisely, Alice computes a public message $\mathsf{S}$ and a secret key $\K$ from her observation $\X^n$; she then transmits $\mathsf{S}$ over the public channel (see Fig. \ref{Figure_SecretKey}). From this message and his own observation $\Y^n$, Bob reconstructs a key $\hat{\K}$. 

Let $\mathcal{K}_n$ and $\mathcal{S}_n$ be the sets of secret keys and public messages respectively. A \emph{secret key rate - public rate pair} $(R_K,R_P)$ is achievable if there exists a sequence of protocols with
$$ \liminf_{n \to \infty} \frac{1}{n} \log \abs{\mathcal{K}_n} \geq R_K, \quad \limsup_{n \to \infty} \frac{1}{n}\log \abs{\mathcal{S}_n} \leq R_P, $$
such that the following properties hold:
\begin{align*}
& \lim_{n \to \infty} \log \abs{\mathcal{K}_n} - \mathbb{H}(\K) = 0 \quad &\text{(uniformity)} \\
&\lim_{n \to \infty} \mathbb{P} \left\{ \K \neq \hat{\K} \right\}=0 \quad &\text{(reliability)} \\
& \lim_{n \to \infty} \mathbb{I}(\K;\mathsf{S},\Z^n)=0\quad & \text{(strong secrecy)}.
\end{align*}
Following \cite{WaOh10}, we denote 
$$\mathcal{R}(\X,\Y,\Z)=\{(R_P,R_K):\; (R_P,R_K) \;\text{is achievable}\}.$$

The optimal trade-off between secret key rate and public rate was derived in \cite{WaOh10}. For the source model (\ref{XYZ-model}), given public rate $R_P$, the secret key rate is upper bounded by
\begin{equation} \label{trade_off_definition}
\!R_K \leq \bar{R}_K(R_P)= \frac{1}{2} \log \left(e^{-2R_P}+\frac{\sigma_2^2}{\sigma_1^2}(1-e^{-2R_P})\right).
\end{equation}
See Appendix \ref{Appendix_tradeoff} for details.

We recall that the secret key capacity of the Gaussian source model (\ref{XYZ-model}) is defined as the maximum achievable secret key rate with unlimited public communication  and is given by 
\begin{align} \label{Cs_definition}
& C_s\!=\!\sup\left\{R_K \; \text{such that}\;\exists R_P \geq 0:\, (R_P,R_K) \in \mathcal{R}(\X,\Y,\Z)\right\} \notag\\
&=
\frac{1}{2}\log\frac{\sigma_2^2}{\sigma_1^2}.
\end{align}

\emph{Additional notation.}
To simplify notation, we define $\hat{\Y}^n=\rho_{xy}\frac{\sigma_x}{\sigma_y} \Y^n$ and $\hat{\Z}^n=\rho_{xz}\frac{\sigma_x}{\sigma_z} \Z^n$, so that
\begin{equation} \label{modified_XYZ-model}
\left\{
\begin{split} 
\X^n&=\hat{\Y}^n+\mathsf{W}_1^n,\\
\X^n&=\hat{\Z}^n+\mathsf{W}_2^n,
\end{split}
\right.
\end{equation}
where $\hat{\Y}^n$ and $\mathsf{W}_1^n$ are independent, and  $\hat{\Z}^n$ and $\mathsf{W}_2^n$ are independent. We denote the variances of $\hat{\Y}^n$ and $\hat{\Z}^n$ by $\hat{\sigma}_y=\rho_{xy}\sigma_x=\sqrt{\sigma_x^2-\sigma_1^2}$ and $\hat{\sigma}_z=\rho_{xz}\sigma_x=\sqrt{\sigma_x^2-\sigma_2^2}$ respectively.

\subsection{Secret key generation protocol} \label{section_protocol}
To define our key generation scheme, we use the lattice partition chain $\Lambda_1/\Lambda_2/\Lambda_3$, where
\begin{itemize}
  \item $\Lambda_1$ is $L^1$ secrecy-good with respect to $\sigma_Q$, and serves as the ``source-code" component of Wyner-Ziv coding;
  \item $\Lambda_2$ is AWGN-good with respect to $\tilde{\sigma}_1=\sqrt{\sigma_1^2+\sigma_Q^2}$, and serves as the ``channel-code" component in Wyner-Ziv coding;
  \item $\Lambda_3$ is $L^1$ secrecy-good with respect to $\tilde{\sigma}_2=\sqrt{\sigma_2^2+\sigma_Q^2}$, and serves as the extractor of randomness.
\end{itemize}
The parameter $\sigma_Q$ controls the quantization rate.\\
The existence of such a chain of lattices will be established in Appendix \ref{existence_section}.

\begin{figure}
\begin{center}
\begin{scriptsize}
\begin{tikzpicture}[scale=0.3,x={(1,0)},y={(0.5,0.5*sqrt(3))}]
\draw[thick,mygreen] (9,0) --(0,9) -- (-9,9)--(-9,0)--(0,-9) -- (9,-9) --cycle ;
\node at (9,0) {\redalert{$\bullet$}}; 
\node at (-9,0) {\redalert{$\circ$}}; 
\node at (0,9) {\redalert{$\bullet$}}; 
\node at (0,-9) {\redalert{$\circ$}}; 
\node at (9,-9) {\redalert{$\circ$}}; 
\node at (-9,9) {\redalert{$\circ$}}; 
\foreach \z/\w in {0/0,3/3,-3/6,-3/-3,3/-6,-6/3,6/-3}
{
\draw[shift={(\z,\w)}] node (0,0) {\redalert{$\bullet$}}; 
\draw[myred,shift={(\z,\w)}, thick] (3,0) --(0,3) -- (-3,3)--(-3,0)--(0,-3) -- (3,-3) --cycle ;
}

\foreach \x/\y in {0/0,1/1,-1/2,-1/-1,1/-2,-2/1,2/-1}
{
\draw[shift={(\x,\y)}] node (0,0) {\myalert{$\bullet$}}; 
\draw[myblue, shift={(\x,\y)}] (1,0) --(0,1) -- (-1,1)--(-1,0)--(0,-1) -- (1,-1) --cycle ;
}
\node at (3,0) {\myalert{$\bullet$}}; 
\node at (-3,0) {\myalert{$\circ$}}; 
\node at (0,3) {\myalert{$\bullet$}}; 
\node at (0,-3) {\myalert{$\circ$}}; 
\node at (3,-3) {\myalert{$\circ$}}; 
\node at (-3,3) {\myalert{$\circ$}};

 \end{tikzpicture}
\end{scriptsize}
 \end{center}
\caption{A schematic representation of the chain of nested lattices $\Lambda_1 \supset \Lambda_2 \supset \Lambda_3$. The fundamental regions of $\Lambda_1$, $\Lambda_2$ and $\Lambda_3$ are pictured in blue, red and green respectively. The quotient groups $\Lambda_1/\Lambda_2$ and $\Lambda_2/\Lambda_3$ are represented by the blue and red points respectively.} 
\label{Fig2}
\end{figure}
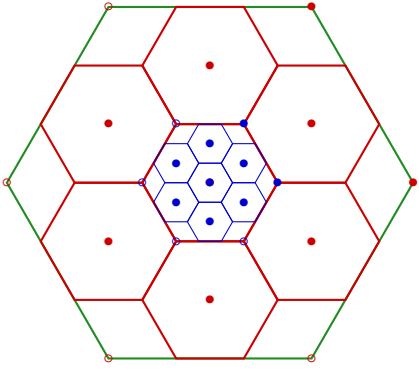

In addition, we assume that $\U$ is a uniform dither over a fundamental region $\mathcal{R}(\Lambda_1)$, which is known by Alice, Bob and Eve\footnote{If Alice and Bob already share a secret source of randomness, there is no need for secret key generation. Hence, Eve should know $\U$ to avoid trivializing the problem.}.

Our protocol is similar to the secret key generation scheme in our previous work \cite{LLB13} with some notable differences due to switching from an $L^{\infty}$ flatness factor criterion to an $L^1$ flatness factor criterion:
\begin{enumerate}
\item[-] As in \cite{LLB13}, the modulo $\mathcal{R}(\Lambda_3)$ operation is used for privacy amplification. Since the the flatness factor $\epsilon^{1}_{\Lambda_3}(\sigma)$ only depends on $f_{\sigma,\Lambda_3}$ which is periodic mod $\Lambda_3$,
nearest-neighbor quantization is not needed 
and 
we can choose any fundamental region $\mathcal{R}(\Lambda_3)$. 
Note that the
mod $\mathcal{R}(\Lambda)$ operation can be performed in polynomial time for many fundamental regions. 
In particular, we can choose the fundamental
parallelepiped. 

\item[-] Nearest-neighbor quantization with respect to the intermediate lattice $\Lambda_2$ is performed for information reconciliation. 
\item[-] As in \cite{LLB13}, quantization with respect to the fine lattice $\Lambda_1$ is performed to obtain a discrete key. However, deterministic quantization is replaced with randomized rounding (using local randomness at Alice's side), which allows to achieve the optimal trade-off between secret key rate and public rate.
Since the $L^1$ flatness factor is only an average condition, dithering is required in order to obtain almost uniform keys. Again, since an $L^1$ flatness factor criterion is used, the dither can be generated uniformly over any fundamental region $\mathcal{R}(\Lambda_1)$. 
\end{enumerate}

\begin{figure}
\begin{small}
\begin{center}
\begin{tikzpicture}[scale=0.3,x={(1,0)},y={(0.5,0.5*sqrt(3))}]
\draw[thick, mygreen] (9,0) --(0,9) -- (-9,9)--(-9,0)--(0,-9) -- (9,-9) --cycle ;
\draw node at (9.5,-10) {\greenalert{$\mathcal{R}(\Lambda_3)$}};

\node at (0,0) {$\bullet$};
\node at (0.5,-1) {$\mb{0}$};
\draw node at (3.5,-4) {\redalert{$\mathcal{V}(\Lambda_2)$}};

\foreach \z/\w in {0/0}
{
\draw[shift={(\z,\w)}] node (0,0) {$\bullet$}; 
\draw[myred,shift={(\z,\w)}, thick] (3,0) --(0,3) -- (-3,3)--(-3,0)--(0,-3) -- (3,-3) --cycle ;
}

\draw[shift={(-4,-7)}] node (0,0) {$\bullet$};
\draw[dashed, thick,shift={(-4,-7)}] (1,0) --(0,1) -- (-1,1)--(-1,0)--(0,-1) -- (1,-1) --cycle ;

\draw[shift={(-2,-8)}] node (0,0) {$\X_Q$}; 

\draw[shift={(-6,-6)}] node (0,0) {$\bullet$};
\draw[shift={(-5.5,-5)}] node (0,0) {$Q_{\Lambda_2}(\X_Q)$}; 
\draw[dashed,shift={(-6,-6)}, thick] (3,0) --(0,3) -- (-3,3)--(-3,0)--(0,-3) -- (3,-3) --cycle ;

\draw[shift={(5,2)}] node (0,0) {$\bullet$}; 
\draw[dashed, thick,shift={(5,2)}] (1,0) --(0,1) -- (-1,1)--(-1,0)--(0,-1) -- (1,-1) --cycle ;
\draw[shift={(7,1)}] node (0,0) {$\bar{\X}_Q$};

\draw[shift={(2,-1)}] node (0,0) {$\myalert{\bullet}$};
\draw[shift={(7.75,-1)}] node (0,0) {\myalert{$\mathsf{S}=\X_Q \Mod \mathcal{V}(\Lambda_2)$}}; 
\draw[myblue, thick,shift={(2,-1)}] (1,0) --(0,1) -- (-1,1)--(-1,0)--(0,-1) -- (1,-1) --cycle ;

\draw[shift={(3,3)}] node (0,0) {$\redalert{\bullet}$};
\draw[shift={(7.5,4)}] node (0,0) { \redalert{$\mathsf{K}=Q_{\Lambda_2}(\X_Q) \Mod \mathcal{R}(\Lambda_3)$}}; 
\draw[myred,shift={(3,3)}, thick] (3,0) --(0,3) -- (-3,3)--(-3,0)--(0,-3) -- (3,-3) --cycle ;

 \end{tikzpicture}
 \end{center}
\end{small}
\caption{A schematic representation of the quantized signal $\X_Q$, the secret key $\K$ and the public message $\S$.} \label{Fig3}
\end{figure}
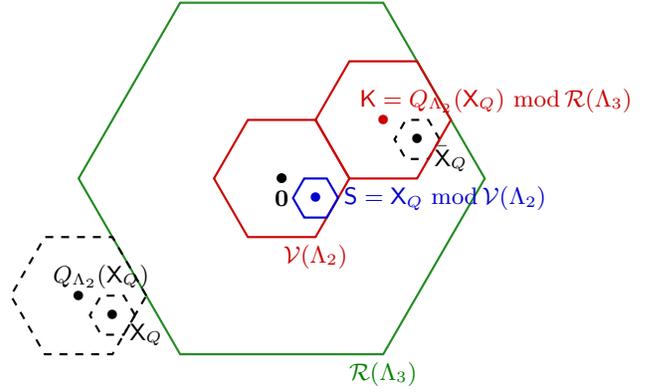

More precisely, the secret key generation proceeds as follows (see Figure \ref{Fig3}):
\begin{itemize}
\item Alice quantizes $\X^n$ to
\begin{equation}
\X_Q=\nint{\X^n+\U}_{\Lambda_1,\sigma_Q}, \label{Alice_quantization}
\end{equation}
according to the randomized rounding operation defined in (\ref{randomized_rounding_definition}). That is, $\X_Q \sim D_{\Lambda_1,\sigma_Q,\mb{x}+\mb{u}}$ if $\X^n=\mb{x}$, $\U=\mb{u}$, or equivalently
\begin{equation} \label{conditional_distribution_x} 
p_{\X_Q|\X^n,\U}(\mb{x}_Q|\mb{x},\mb{u})=\frac{f_{\sigma_Q}(\mb{x}_Q-\mb{x}-\mb{u})}{f_{\sigma_Q}(\Lambda_1-\mb{x}-\mb{u})}.
\end{equation}
Alice then computes the public message $\mathsf{S} \in \mathcal{S}=\Lambda_1/\Lambda_2$ and the key $\K \in \mathcal{K}=\Lambda_2/\Lambda_3$ as follows: 
\begin{align*}
&\mathsf{S}=\X_Q \Mod \mathcal{V}(\Lambda_2),\\
& \K=Q_{\Lambda_2}(\X_Q) \Mod \mathcal{R}(\Lambda_3),
\end{align*}
and transmits $\mathsf{S}$ to Bob over the public channel. 
\item Upon receiving $\mathsf{S}$, Bob reconstructs
$$\hat{\X}_Q=\mathsf{S}+Q_{\Lambda_2}\left(\rho_{xy} \frac{\sigma_x}{\sigma_y} \Y^n+\U -\mathsf{S}\right).$$
He then computes his version of the key:
$$\hat{\K}=Q_{\Lambda_2}(\hat{\X}_Q) \Mod \mathcal{R}(\Lambda_3).$$
\end{itemize}

Let $\bar{\X}_Q=\X_Q \Mod \mathcal{R}(\Lambda_3) \in \Lambda_1/\Lambda_3$, where the quotient $\Lambda_1/\Lambda_3$ is identified with the set of coset representatives $\Lambda_1 \cap \mathcal{R}(\Lambda_3)$. 
By definition, $\bar{\X}_Q=\mathsf{S}+\K$.
Note that $\K$ and $\mathsf{S}$ are both functions of $\bar{\X}_Q$: 
\begin{align} \label{K_function}
& \K=Q_{\Lambda_2}(\X_Q) \Mod \mathcal{R}(\Lambda_3)
\notag\\
&\stackrel{(a)}{=}Q_{\Lambda_2}(\X_Q \Mod \mathcal{R}(\Lambda_3))\Mod \mathcal{R}(\Lambda_3) \notag \\
&=Q_{\Lambda_2}(\bar{\X}_Q) \Mod \mathcal{R}(\Lambda_3)=f(\bar{\X}_Q).
\end{align}
where $(a)$ follows from equation (\ref{NG_35}). Similarly,
\begin{align} \label{S_function}
& \bar{\X}_Q \Mod \Lambda_2=\bar{\X}_Q-Q_{\Lambda_2}(\bar{\X}_Q) \notag \\
&=
\X_Q-Q_{\mathcal{R}(\Lambda_3)}(\X_Q)-Q_{\Lambda_2}(\X_Q-Q_{\mathcal{R}(\Lambda_3)}(\X_Q))\notag \\
&=
\X_Q-Q_{\Lambda_2}(\X_Q) =\X_Q \Mod \Lambda_2=\mathsf{S}=g(\bar{\X}_Q).
\end{align}

\begin{rem} \label{bijection}
Because of the previous relations, we can conclude that there exists a bijection $(f,g): \Lambda_1/\Lambda_3 \to \Lambda_1/\Lambda_2 \times \Lambda_2/\Lambda_3$ that sends $\bar{\X}_Q$ into the corresponding pair $(\S,\K)$. 
\end{rem}

We now state the main result of the paper, which will be
proven in the following sections:
\begin{theorem} \label{main_theorem}
For the Gaussian source model (\ref{XYZ-model}), there exists a sequence of nested lattices $\Lambda_3^{(n)} \subset \Lambda_2^{(n)} \subset \Lambda_1^{(n)}$ such that for any public rate $R_P>0$, the previous secret key generation protocol asymptotically achieves the optimal secret key rate $\bar{R}_K(R_P)$ in (\ref{trade_off_definition}). In particular, any
secret key rate $R_K<C_s=\frac{1}{2}\log\frac{\sigma_2^2}{\sigma_1^2}$ is achievable.  
\end{theorem}

\subsection{Properties of randomized rounding and discrete Gaussians} \label{section_randomized_rounding}
Before proceeding to prove Theorem \ref{main_theorem}, we need some preliminary results about the properties of the randomized quantization in equation (\ref{Alice_quantization}).
It was shown in \cite{Peikert} that when 
$\X^n$ is i.i.d. Gaussian with variance $\sigma^2$, the randomly rounded variable $\nint{\X^n}_{\Lambda, \sigma_Q}$ is close in variational distance to the discrete Gaussian $D_{\Lambda,\tilde{\sigma}}$, where 
$\tilde{\sigma}^2=\sigma^2+\sigma_Q^2$, provided that the $L^{\infty}$ flatness factor $\epsilon_{\Lambda}(\sigma_Q)$ is small:

\begin{prop}[Adapted from Theorem 3.1 of \cite{Peikert}]
Let $\X^n \sim \mathcal{N}(0,\sigma^2I_n)$ and $\boldsymbol\mu \in \R^n$, and consider $\X_Q=\nint{\X^n+\boldsymbol\mu}_{\Lambda,\sigma_Q}$. If $\epsilon_{\Lambda}(\sigma_Q)<1/2$, then
$$\V(p_{\X_Q},D_{\Lambda,\tilde{\sigma},\boldsymbol\mu)}) \leq 4 \epsilon_{\Lambda}(\sigma_Q),$$
where $\tilde{\sigma}^2=\sigma^2+\sigma_Q^2$.
\end{prop}

In the following, we prove a partial generalization of this result under an $L^1$ flatness factor condition, for randomized rounding with uniform dithering, which may be of independent interest. 

\begin{lem} \label{lemma_generalized_Peikert}
Given a Gaussian random vector $\X^n \sim \mathcal{N}(0,\sigma^2I_n)$, a dither $\mathsf{U} \sim \mathcal{U}_{\mathcal{R}}$ uniform over a fundamental region $\mathcal{R}$ of the lattice $\Lambda$ and independent of $\X^n$, and a constant $\boldsymbol\mu \in \R^n$, let $\X_Q=\nint{\X^n+\U+\boldsymbol\mu}_{\Lambda, \sigma_Q}$. Then 
$$\mathbb{E}_{\U}\left[\V\left(p_{\X_Q|\U},D_{\Lambda,\tilde{\sigma},\U+\boldsymbol\mu}\right)\right] \leq 2 \epsilon_{\Lambda}^{1}(\sigma_Q).$$ 
\end{lem}

In order to prove Lemma \ref{lemma_generalized_Peikert}, we need the following intermediate Lemma.

\begin{lem} \label{intermediate_lemma}
Suppose that $\tilde{\sigma}^2=\sigma^2+\sigma_Q^2$, and let $\mathcal{R}$ be a fundamental region of $\Lambda$. Then the following inequality holds:
\begin{align*}
&\!\!\sum_{\mb{x}_Q \in \Lambda} \!\int_{\mathcal{R}} \!\abs{\int_{\R^n}\!\!\! \frac{f_{\sigma}(\mb{x}\!-\!\boldsymbol\mu)f_{\sigma_Q}\!(\mb{x}_Q\!-\!\mb{x}\!-\!\mb{u})}{V(\Lambda) f_{\sigma_Q}\!(\Lambda-\mb{x}-\mb{u})}d\mb{x}\!-\!f_{\tilde{\sigma}}(\mb{x}_Q\!-\!\mb{u}\!-\!\boldsymbol\mu)}\!d\mb{u}\\ &\leq \epsilon^1_{\Lambda}(\sigma_Q).
\end{align*}
\end{lem}
\begin{IEEEproof}[Proof of Lemma \ref{intermediate_lemma}]
By Lemma \ref{Fact1}, 
\begin{equation} \label{eq_Lemma1}
f_{\sigma_Q}(\mb{x}_Q-\mb{x}-\mb{u})f_{\sigma}(\mb{x}-\boldsymbol\mu)=f_{\tilde{\sigma}}(\mb{x}_Q-\mb{u}-\boldsymbol\mu)f_{\bar{\sigma}}(\mb{x}-\bar{\mb{c}}),
\end{equation}
where $\frac{1}{\bar{\sigma}^2}=\frac{1}{\sigma^2}+\frac{1}{\sigma_Q^2}$ and $\bar{\mb{c}}=\frac{\bar{\sigma}^2}{\sigma_Q^2}(\mb{x}_Q-\mb{u})+\frac{\bar{\sigma}^2}{\sigma^2}\boldsymbol\mu$. Then we can write
{\allowdisplaybreaks
\begin{align*}
&\!\!\sum_{\mb{x}_Q \in \Lambda} \!\int_{\mathcal{R}}\! \abs{\int_{\R^n} \!\!\!\frac{f_{\sigma}(\mb{x}\!-\!\boldsymbol\mu)f_{\sigma_Q}\!(\mb{x}_Q\!-\!\mb{x}\!-\!\mb{u})}{V(\Lambda) f_{\sigma_Q}(\Lambda-\mb{x}\!-\!\mb{u})}d\mb{x}\!-\!f_{\tilde{\sigma}}(\mb{x}_Q\!-\!\mb{u}\!-\!\boldsymbol\mu)}\!d\mb{u}\\
&\stackrel{(a)}{=}
\begin{multlined}[t]\sum_{\mb{x}_Q \in \Lambda} \int_{\mathcal{R}} 
\bigg|\int_{\R^n} \frac{f_{\sigma}(\mb{x}-\boldsymbol\mu)f_{\sigma_Q}(\mb{x}_Q-\mb{x}-\mb{u})}{V(\Lambda) f_{\sigma_Q}(\Lambda-\mb{x}-\mb{u})}d\mb{x}\\-f_{\tilde{\sigma}}(\mb{x}_Q-\mb{u}-\boldsymbol\mu)\int_{\R^n} f_{\bar{\sigma}}(\mb{x}-\bar{\mb{c}})d\mb{x}
\bigg|
d\mb{u}
\end{multlined}\\
&\stackrel{(b)}{=}
\begin{multlined}[t]
\sum_{\mb{x}_Q \in \Lambda} \int_{\mathcal{R}} 
\bigg|
\int_{\R^n} \frac{f_{\sigma}(\mb{x}-\boldsymbol\mu)f_{\sigma_Q}(\mb{x}_Q-\mb{x}-\mb{u})}{V(\Lambda) f_{\sigma_Q}(\Lambda-\mb{x}-\mb{u})}d\mb{x}\\
-\int_{\R^n} f_{\sigma_Q}(\mb{x}_Q-\mb{x}-\mb{u})f_{\sigma}(\mb{x}-\boldsymbol\mu)d\mb{x}
\bigg|
d\mb{u} 
\end{multlined}\\
& \leq \begin{multlined}[t][0.9\linewidth]
\int_{\mathcal{R}} \!\int_{\R^n}\!  \!\frac{\sum_{\mb{x}_Q \in \Lambda}f_{\sigma}(\mb{x}\!-\!\boldsymbol\mu)f_{\sigma_Q}(\mb{x}_Q\!-\!\mb{x}\!-\!\mb{u})}{f_{\sigma_Q}(\Lambda\!-\!\mb{x}\!-\!\mb{u})}\cdot\\
\cdot \abs{ 
\frac{1}{V(\Lambda)} -f_{\sigma_Q}(\Lambda\!-\!\mb{x}\!-\!\mb{u})
}
d\mb{x}d\mb{u} \end{multlined}\\
& = \int_{\R^n}  f_{\sigma}(\mb{x}-\boldsymbol\mu) \int_{\mathcal{R}} \abs{ \frac{1}{V(\Lambda)} -f_{\Lambda,\sigma_Q}(\mb{x}+\mb{u})}d\mb{u} d\mb{x}\\
& =\int_{\R^n}  f_{\sigma}(\mb{x}-\boldsymbol\mu) \int_{\mathcal{R}} \abs{ \frac{1}{V(\Lambda)} -f_{\Lambda,\sigma_Q}(\mb{u})}d\mb{u} d\mb{x}
=\epsilon^1_{\Lambda}(\sigma_Q),
\end{align*}
}%
where $(a)$ follows from the fact that $\int_{\R^n} f_{\bar{\sigma}}(\mb{x}-\bar{\mb{c}})d\mb{x}=1$, and $(b)$ follows from (\ref{eq_Lemma1}). 
\end{IEEEproof}
\medskip \par
\begin{IEEEproof}[Proof of Lemma \ref{lemma_generalized_Peikert}]
We have
{\allowdisplaybreaks
\begin{align}
&\mathbb{E}_{\U}\left[\V\left(p_{\X_Q|\U},D_{\Lambda,\tilde{\sigma},\U+\boldsymbol\mu}\right)\right]\notag \\
&=
\sum_{\mb{x}_Q \in \Lambda} \int_{\mathcal{R}} \frac{1}{V(\Lambda)} 
\bigg|p_{\X_Q|\U}(\mb{x}_Q|\mb{u})
- \frac{f_{\tilde{\sigma}}(\mb{x}_Q-\mb{u}-\boldsymbol\mu)}{f_{\tilde{\sigma}}(\Lambda-\mb{u}-\boldsymbol\mu)}
\bigg|
d\mb{u} 
\notag\\
&\!\stackrel{(a)}{\leq}\!\! \!\!\sum_{\mb{x}_Q \in \Lambda} \!\int_{\mathcal{R}}\! 
\abs{
p_{\X_Q|\U}(\mb{x}_Q|\mb{u})
\!-\!\!
f_{\tilde{\sigma}}(\mb{x}_Q\!-\!\mb{u}\!-\!\boldsymbol\mu)
}\!
d\mb{u} \notag\\
& + \sum_{\mb{x}_Q \in \Lambda} \int_{\mathcal{R}}  \abs{f_{\tilde{\sigma}}(\mb{x}_Q\!-\!\mb{u}\!-\!\boldsymbol\mu) \!-\! \frac{f_{\tilde{\sigma}}(\mb{x}_Q\!-\!\mb{u}\!-\!\boldsymbol\mu)}{V(\Lambda)f_{\tilde{\sigma}}(\Lambda\!-\!\mb{u}\!-\!\boldsymbol\mu)}}d\mb{u} \label{B2},
\end{align}
}%
where $(a)$ follows from the triangle inequality. 

We note that
\begin{align*}
&p_{\X_Q|\U}(\mb{x}_Q|\mb{u})=
\int_{\R^n} 
p_{\X_Q|\X^n,\U}(\mb{x}_Q|\mb{x},\mb{u})p_{\X^n}(\mb{x})d\mb{x}\\
&=\int_{\R^n} \frac{f_{\sigma}(\mb{x})f_{\sigma_Q}(\mb{x}_Q-\mb{x}-\mb{u}-\boldsymbol\mu)}{f_{\sigma_Q}(\Lambda-\mb{x}-\mb{u}-\boldsymbol\mu)}d\mb{x}\\
&=\int_{\R^n}\frac{f_{\sigma}(\mb{x}-\boldsymbol\mu)f_{\sigma_Q}(\mb{x}_Q-\mb{x}-\mb{u})}{f_{\sigma_Q}(\Lambda-\mb{x}-\mb{u})}d\mb{x}.
\end{align*}%
Thus, the first term in (\ref{B2}) is bounded by $\epsilon^1_{\Lambda}(\sigma_Q)$ because of Lemma \ref{intermediate_lemma}.  The second term in (\ref{B2}) is equal to
\begin{align*}
&\sum_{\mb{x}_Q \in \Lambda} \int_{\mathcal{R}} \frac{f_{\tilde{\sigma}}(\mb{x}_Q-\mb{u}-\boldsymbol\mu)}{f_{\tilde{\sigma}}(\Lambda-\mb{u}-\boldsymbol\mu)} \abs{f_{\tilde{\sigma}}(\Lambda-\mb{u}-\boldsymbol\mu)-\frac{1}{V(\Lambda)}}d\mb{u}\\
&=\int_{\mathcal{R}}\abs{f_{\tilde{\sigma}}(\Lambda-\mb{u}-\boldsymbol\mu)-\frac{1}{V(\Lambda)}}d\mb{u}=\epsilon^1_{\Lambda}(\tilde{\sigma}) \stackrel{(b)}{\leq}  \epsilon^{1}_{\Lambda}(\sigma_Q),
\end{align*}%
where $(b)$ follows from Lemma \ref{L1_monotonic}.  
\end{IEEEproof}
Another useful property of discrete Gaussian distributions is that a sample $D_{\Lambda,\sigma,\mb{c}}$ is distributed almost uniformly modulo a sublattice $\Lambda' \subset \Lambda$ provided that $\epsilon_{\Lambda'}(\sigma)$ is small \cite[Corollary 2.8]{GPV}:
\begin{prop}
Let $\Lambda'\subset \Lambda$. Then if $\epsilon_{\Lambda'}(\sigma)<1$,  
$$\norm{D_{\Lambda,\sigma,\mb{c}}\Mod \Lambda'-\mathcal{U}_{\Lambda/\Lambda'}}_{\infty} \leq 4 \epsilon_{\Lambda'}(\sigma)$$
\end{prop}
In the statement above, with slight abuse of notation, $D_{\Lambda,\sigma,\mb{c}}\Mod \Lambda'$ denotes the probability density of the random variable $\X_D \Mod \Lambda'$, where $\X_D \sim D_{\Lambda,\sigma,\mb{c}}$. 

We can partially generalize this statement in an average sense under an $L^1$-flatness factor condition:
\begin{lem} \label{lemma_generalized_GPV}
Let $\Lambda'\subset \Lambda$. Then 
$$\mathbb{E}_{\U}\left[\V\left(D_{\Lambda, \sigma, \U} \Mod \Lambda', \mathcal{U}_{\Lambda/\Lambda'}\right)\right] \leq 2 \epsilon^1_{\Lambda'}(\sigma)$$
\end{lem}
\begin{IEEEproof}
Given two fundamental regions $\mathcal{R}(\Lambda)$, $\mathcal{R}(\Lambda')$, we can write
\begin{align}
&\mathbb{E}_{\U}\left[\V\left(D_{\Lambda, \sigma, \U} \Mod \Lambda', \mathcal{U}_{\Lambda/\Lambda'}\right)\right] \notag\\
&= \int_{\mathcal{R}(\Lambda)} \frac{1}{V(\Lambda)} \sum_{\smash{\tilde{\lambda} \in \Lambda/\Lambda'}} \abs{\sum_{\lambda'\in \Lambda'} \frac{f_{\sigma,\mb{u}}(\tilde{\lambda}+\lambda')}{f_{\sigma,\Lambda}(\mb{u})}-\frac{V(\Lambda)}{V(\Lambda')}}d\mb{u} \notag\\
&=\int_{\mathcal{R}(\Lambda)}  \sum_{\smash{\tilde{\lambda} \in \Lambda/\Lambda'}} \abs{\sum_{\lambda'\in \Lambda'} \frac{f_{\sigma,\mb{u}}(\tilde{\lambda}+\lambda')}{f_{\sigma,\Lambda}(\mb{u})V(\Lambda)}-\frac{1}{V(\Lambda')}}d\mb{u} \notag \\
& \leq \int_{\mathcal{R}(\Lambda)}  \sum_{\smash{\tilde{\lambda} \in \Lambda/\Lambda'}} \abs{\sum_{\lambda'\in \Lambda'} \frac{f_{\sigma,\mb{u}}(\tilde{\lambda}+\lambda')}{f_{\sigma,\Lambda}(\mb{u})V(\Lambda)}-f_{\sigma,\Lambda'}(\mb{u}+\tilde{\lambda})}d\mb{u} \notag \\
&+\int_{\mathcal{R}(\Lambda)}  \sum_{\smash{\tilde{\lambda} \in \Lambda/\Lambda'}} \abs{f_{\sigma,\Lambda'}(\mb{u}+\tilde{\lambda})-\frac{1}{V(\Lambda')}}d\mb{u} \label{s}
\end{align}
by the triangle inequality.\\
The first term in (\ref{s}) can be rewritten as follows:
\begin{align*}
& \int_{\mathcal{R}(\Lambda)}  \sum_{\smash{\tilde{\lambda} \in \Lambda/\Lambda'}} \abs{\sum_{\lambda'\in \Lambda'} \frac{f_{\sigma,\mb{u}}(\tilde{\lambda}+\lambda')}{f_{\sigma,\Lambda}(\mb{u})V(\Lambda)}-\sum_{\lambda'\in \Lambda'}f_{\sigma,\mb{u}}(\tilde{\lambda}+\lambda')}d\mb{u}\\
&\leq \int_{\mathcal{R}(\Lambda)} \sum_{\smash{\tilde{\lambda} \in \Lambda/\Lambda'}} \sum_{\lambda'\in \Lambda'} \frac{f_{\sigma,\mb{u}}(\tilde{\lambda}+\lambda')}{f_{\sigma, \Lambda}(\mb{u})}\abs{\frac{1}{V(\Lambda)}-f_{\sigma,\Lambda}(\mb{u})} d\mb{u} \\
&=\int_{\mathcal{R}(\Lambda)} \sum_{\lambda \in \Lambda} \frac{f_{\sigma,\mb{u}}(\lambda)}{f_{\sigma,\Lambda}(\mb{u})}\abs{\frac{1}{V(\Lambda)}-f_{\sigma,\Lambda}(\mb{u})} d\mb{u}\\
&=\int_{\mathcal{R}(\Lambda)}\abs{\frac{1}{V(\Lambda)}-f_{\sigma,\Lambda}(\mb{u})} d\mb{u}=\epsilon^1_{\Lambda}(\sigma) \leq \epsilon^1_{\Lambda'}(\sigma) 
\end{align*}
by Remark \ref{remark_nested_lattices}.\\
Setting $\mb{v}=\mb{u}+\tilde{\lambda} \Mod \Lambda'$, the second term is equal to 
\begin{align}
\int_{\mathcal{R}(\Lambda')} \abs{f_{\sigma,\Lambda'}(\mb{v})-\frac{1}{V(\Lambda')}} d\mb{v}=\epsilon^1_{\Lambda'}(\sigma). \tag*{\IEEEQED}
\end{align}
\let\IEEEQED\relax%
\end{IEEEproof}
From Lemma \ref{lemma_generalized_Peikert} and Lemma \ref{lemma_generalized_GPV}, we can immediately deduce the following:
\begin{corol} \label{corollary_randomized_rounding}
Consider two nested lattices $\Lambda'\subset \Lambda$. Given a Gaussian random vector $\X^n \sim \mathcal{N}(0,\sigma^2I_n)$, a dither $\mathsf{U} \sim \mathcal{U}_{\mathcal{R}(\Lambda)}$ uniform over a fundamental region $\mathcal{R}(\Lambda)$ and independent of $\X^n$, and a constant $\boldsymbol\mu \in \R^n$, let $\X_Q=\nint{\X^n+\U+\boldsymbol\mu}_{\Lambda, \sigma_Q}$.
Then 
$$\mathbb{E}_{\U}\left[\V\left(p_{\X_Q \Mod \Lambda'|\U},\mathcal{U}_{\Lambda/\Lambda'}\right)\right] \leq 2 \epsilon^1_{\Lambda}(\sigma_Q)+2\epsilon^1_{\Lambda'}(\tilde{\sigma}),$$
where $\tilde{\sigma}^2=\sigma^2+\sigma_Q^2$.
\end{corol}
\begin{IEEEproof}
We have
{\allowdisplaybreaks
\begin{align*}
&\mathbb{E}_{\U}\left[\V\left(p_{\X_Q\Mod \Lambda'|\U} ,\mathcal{U}_{\Lambda/\Lambda'}\right)\right] \\
& \stackrel{(a)}{\leq} \mathbb{E}_{\U}\left[\V\left(p_{\X_Q\Mod \Lambda'|\U} ,D_{\Lambda,\tilde{\sigma},\U}\Mod \Lambda'\right)\right)] \\
&+
\mathbb{E}_{\U}\left[\V\left(D_{\Lambda,\tilde{\sigma},\U}\Mod \Lambda',\mathcal{U}_{\Lambda/\Lambda'}\right)\right] \\
&\stackrel{(b)}{\leq} \mathbb{E}_{\U}\left[\V\left(p_{\X_Q|\U},D_{\Lambda,\tilde{\sigma},\U}\right)\right)]+2\epsilon^1_{\Lambda'}(\tilde{\sigma}) \\
&\leq 2\epsilon^1_{\Lambda}(\sigma_Q)+2\epsilon^1_{\Lambda'}(\tilde{\sigma}) 
\end{align*}
}%
where (a) follows from the triangle inequality, (b) follows from the data processing inequality for the variational distance and Lemma \ref{lemma_generalized_GPV}, and (c) follows from Lemma \ref{lemma_generalized_Peikert}. 
\end{IEEEproof}

\subsection{Reliability}
\label{section_reliability}
We want to show that the error probability $P_e=\protect{\mathbb{P}\{\K \neq \hat{\K}\}} \to 0$ as $n \to \infty$. Note that $\K=\hat{\K}$ if $\hat{\X}_Q=\X_Q$. Since $\X_Q=\mathsf{S}+Q_{\Lambda_2}(\X_Q)$, we have
$$\hat{\X}_Q=\X_Q \;\Leftrightarrow\; Q_{\Lambda_2}(\hat{\Y}^n+\U -\mathsf{S})=Q_{\Lambda_2}(\X_Q).$$
Observe that 
\begin{align*}
&Q_{\Lambda_2}(\hat{\Y}^n+\U  -\mathsf{S})=Q_{\Lambda_2}\left( \hat{\Y}^n+\U -\X_Q +Q_{\Lambda_2}(\X_Q)\right)\\
&=Q_{\Lambda_2}( \hat{\Y}^n+\U -\X_Q) + Q_{\Lambda_2}(\X_Q).
\end{align*}
Therefore 
\begin{align}  
&\hat{\X}_Q=\X_Q \;\Leftrightarrow\;Q_{\Lambda_2}(\hat{\Y}^n +\U -\X_Q)=0 \notag \\
&\; \Leftrightarrow \; \hat{\Y}^n \in \X_Q-\U + \mathcal{V}(\Lambda_2).
\label{reliability_condition}
\end{align}
The error probability is bounded by
\begin{align*}
&P_e \leq \mathbb{P}\{\hat{\X}_Q \neq \X_Q\}\\
&=\mathbb{E}_{\X^n\hat{\Y}^n\U}\!\left[
\mathbb{P}\{ \hat{\X}_Q\! \neq\! \X_Q | \hat{\Y}^n, \X^n, \U\} \right]\\
&=\!\mathbb{E}_{\X^n\hat{\Y}^n\U}\Bigg[
\!\sum_{\mb{x}_Q \in \Lambda_1}\!\! p_{\X_Q|\X^n\U}(\mb{x}_Q) \mathbb{P}\{ \hat{\X}_Q \!\neq\! \mb{x}_Q | \hat{\Y}^n,\U, \X_Q\!=\!\mb{x}_Q\}\Bigg] \end{align*}
In the last step we have used the Markov chain $\X^n - (\hat{\Y}^n, \X_Q,\U) - \hat{\X}_Q$.
Replacing the expression for the conditional distribution in equation (\ref{conditional_distribution_x}), we obtain
{\allowdisplaybreaks
\begin{align}
& \begin{multlined}[t][0.9\linewidth] P_e  \leq  \sum_{\mb{x}_Q \in \Lambda_1}\bigg( \int_{\R^n} \! \int_{\mathcal{R}(\Lambda_1)}
\mathds{1}_{\{\mb{y} \notin \mb{x}_Q-\mb{u}+\mathcal{V}(\Lambda_2)\}} \cdot \\
\cdot 
\int_{\R^n}\frac{f_{\sigma_Q}(\mb{x}_Q-\mb{x}-\mb{u})}{f_{\sigma_Q}(\Lambda_1-\mb{x}-\mb{u})}  \frac{p_{\X^n|\hat{\Y}^n}(\mb{x}|\mb{y})  p_{\hat{\Y}^n}(\mb{y})}{V(\Lambda_1)} d\mb{x} 
d\mb{u}d\mb{y}\bigg)\end{multlined}\notag\\
& \begin{multlined}[t][0.9\linewidth]   =  \sum_{\mb{x}_Q \in \Lambda_1}\bigg( \int_{\R^n} \! \int_{\mathcal{R}(\Lambda_1)}
\mathds{1}_{\{\mb{y} \notin \mb{x}_Q-\mb{u}+\mathcal{V}(\Lambda_2)\}} f_{\hat{\sigma}_y}(\mb{y})
\cdot \\
\cdot 
\int_{\R^n}\frac{f_{\sigma_Q}(\mb{x}_Q-\mb{x}-\mb{u})}{f_{\sigma_Q}(\Lambda_1-\mb{x}-\mb{u})}  \frac{f_{\sigma_1}(\mb{x}-\mb{y})}{V(\Lambda_1)}d\mb{x} 
d\mb{u}d\mb{y}\bigg)\end{multlined}\notag\\
& \stackrel{(a)}{\leq}
\sum_{\mb{x}_Q \in \Lambda_1}\!\! \bigg(
\int_{\R^n}  \int_{\mathcal{R}(\Lambda_1)}
\mathds{1}_{\{\mb{y} \notin \mb{x}_Q-\mb{u}+\mathcal{V}(\Lambda_2)\}} f_{\hat{\sigma}_y}(\mb{y}) 
\cdot \notag\\
& \cdot
\bigg|
\int_{\R^n}\!\!\frac{f_{\sigma_Q}(\mb{x}_Q\!-\!\mb{x}\!-\!\mb{u})f_{\sigma_1}(\mb{x}\!-\!\mb{y})}{f_{\sigma_Q}(\Lambda_1\!-\!\mb{x}\!-\!\mb{u})V(\Lambda_1)}d\mb{x} \!-\!f_{\tilde{\sigma}_1}(\mb{x}_Q\!-\!\mb{u}\!-\!\mb{y})
\bigg|
 d\mb{u}d\mb{y}\!\bigg) 
\notag
\\
&+\!\!\!\sum_{\mb{x}_Q \in \Lambda_1}\!\! \int_{\R^n}\!  \int_{\mathcal{R}(\Lambda_1)}\!\!\! f_{\tilde{\sigma}_1}\!(\mb{x}_Q\!-\!\mb{u}\!-\!\mb{y})\mathds{1}_{\!\{\mb{y} \notin \mb{x}_Q\!-\!\mb{u}\!+\!\mathcal{V}(\Lambda_2)\}}\! f_{\hat{\sigma}_y}\!(\mb{y}) d\mb{u} d\mb{y}   \label{b}
\end{align}
}%
where $(a)$ follows from the triangle inequality.\\
The first term of (\ref{b}) is upper bounded by
$\epsilon^1_{\Lambda_1}(\sigma_Q)$ 
using Lemma \ref{intermediate_lemma}. This tends to $0$ provided that $\Lambda_1$ is $L^1$ secrecy-good and 
\begin{equation} \label{condition_Lambda1}
\frac{V(\Lambda_1)^{2/n}}{\sigma_Q^2}< 2 \pi e.
\end{equation} 
With the change of variables $\mb{y}'=\mb{y}-\mb{x}_Q+\mb{u}$, the second term of (\ref{b}) can be rewritten as
{\allowdisplaybreaks
\begin{align*}
&\sum_{\mb{x}_Q \in \Lambda_1} \int_{\mathcal{R}(\Lambda_1)} \int_{\R^n} f_{\tilde{\sigma}_1}(\mb{y}') \mathds{1}_{\{\mb{y}'\notin \mathcal{V}(\Lambda_2)\}}f_{\hat{\sigma}_y}(\mb{y}'\!+\!\mb{x}_Q\!-\!\mb{u}) d\mb{y}'d\mb{u} \\
&=\sum_{\mb{x}_Q \in \Lambda_1} \int_{\mathcal{R}(\Lambda_1)} \int_{\R^n\setminus \mathcal{V}(\Lambda_2)} f_{\tilde{\sigma}_1}(\mb{y}') f_{\hat{\sigma}_y}(\mb{y}'+\mb{x}_Q-\mb{u}) d\mb{y}'d\mb{u}\\
&=\int_{\R^n\setminus \mathcal{V}(\Lambda_2)} f_{\tilde{\sigma}_1}(\mb{y}') \int_{\mathcal{R}(\Lambda_1)} f_{\hat{\sigma}_y, \Lambda_1}(\mb{y}'-\mb{u})d\mb{u} d\mb{y}'\\
&\stackrel{(b)}{=}\int_{\R^n\setminus \mathcal{V}(\Lambda_2)} f_{\tilde{\sigma}_1}(\mb{y}') d\mb{y}'
\end{align*}
}%
where $(b)$ holds since $\int_{\mathcal{R}(\Lambda_1)} f_{\hat{\sigma}_y, \Lambda_1}(\mb{y}'-\mb{u})d\mb{u}=1$.
This tends to $0$ provided that $\Lambda_2$ is AWGN-good and
\begin{equation} \label{condition_Lambda2}
\frac{V(\Lambda_2)^{2/n}}{\tilde{\sigma}_1^2}>2\pi e.
\end{equation} 
\subsection{Uniformity} \label{uniformity_section}

We want to show that the key is asymptotically uniform when $n \to \infty$. 
Let $\tilde{\sigma}_x^2=\sigma_x^2+\sigma_Q^2$. First, we bound the $L^1$ distance between $p_{\bar{\X}_Q}$ and the uniform distribution over $\Lambda_1/\Lambda_3$: 
\begin{align}
&\mathbb{V}(p_{\bar{\X}_Q},\mathcal{U}_{\Lambda_1/\Lambda_3}) \notag\\
&\stackrel{(a)}{\leq}\mathbb{E}_{\U}\left[\V\left(p_{\bar{\X}_Q|\U} ,\mathcal{U}_{\Lambda_1/\Lambda_3}\right)\right] \notag\\
&\stackrel{(b)}{\leq} 2 \epsilon^1_{\Lambda_1}(\sigma_Q)+2\epsilon^1_{\Lambda_3}(\tilde{\sigma}_x) \notag\\
&\stackrel{(c)}{\leq} 2 \epsilon^1_{\Lambda_1}(\sigma_Q)+2\epsilon^1_{\Lambda_3}(\tilde{\sigma}_2)
\label{bound_2}
\end{align}
where (a) follows from Lemma 7 in \cite{Bloch_Laneman},
(b) follows from Corollary \ref{corollary_randomized_rounding} and (c) follows from Lemma \ref{L1_monotonic}, since $\tilde{\sigma}_2^2=\sigma_2^2+\sigma_Q^2 \leq \sigma_x^2+\sigma_Q^2=\tilde{\sigma}_x^2$.\\ 
The term (\ref{bound_2}) vanishes as 
$o\left(\frac{1}{n}\right)$ if both 
$\Lambda_1$ and
$\Lambda_3$ are $L^1$-secrecy good and satisfy the volume conditions (\ref{condition_Lambda1}) and 
\begin{equation} \label{condition_Lambda3}
\frac{V(\Lambda_3)^{2/n}}{\tilde{\sigma}_2^2}<2\pi e.
\end{equation}
We note that actually a slightly tighter bound than (\ref{bound_2}) holds, where the coefficient $2$ is replaced by $1$.\footnote{This bound can be obtained using Lemma \ref{intermediate_lemma}, see the preprint version of this work \cite{secret_key_preprint}. Here, we prefer to state Lemmas \ref{lemma_generalized_Peikert} and \ref{lemma_generalized_GPV}, which shorten the proof, have a clearer operational meaning and can be of independent interest.}\\
We now show that the distribution of the key is close to the uniform distribution $\mathcal{U}_{\mathcal{K}}$ over $\mathcal{K}=\Lambda_2/\Lambda_3$:
{\allowdisplaybreaks
\begin{align*}
&\mathbb{V}(p_{\K},\mathcal{U}_{\mathcal{K}})=\sum_{k \in \mathcal{K}} \abs{ p_{\K}(k)-\frac{V(\Lambda_2)}{V(\Lambda_3)}} \\
&=
\sum_{k \in \mathcal{K}} \abs{\sum_{s \in \mathcal{S}} p_{\bar{\X}_Q}(s + k) -\sum_{s \in \mathcal{S}}\frac{V(\Lambda_1)}{V(\Lambda_3)} } \\
&\leq \sum_{k \in \mathcal{K}} \sum_{s \in \mathcal{S}} \abs{p_{\bar{\X}_Q}(s + k) - \frac{V(\Lambda_1)}{V(\Lambda_3)}}\\ &=\sum_{\bar{\mb{x}}_Q \in \Lambda_1/\Lambda_3} \abs{p_{\bar{\X}_Q}(\bar{\mb{x}}_Q)-\frac{V(\Lambda_1)}{V(\Lambda_3)}}=\mathbb{V}(p_{\bar{\X}_Q},\mathcal{U}_{\Lambda_1/\Lambda_3})
\end{align*}
}%
which vanishes as $o\left(\frac{1}{n}\right)$ as shown previously.
Using \cite[Lemma 2.7]{Csiszar_Korner}, we have that if $\V(p_{\K},\mathcal{U}_{\mathcal{K}}) \leq \frac{1}{2}$,
{\allowdisplaybreaks
\begin{align*}
&\abs{\mathbb{H}(p_{\K})-\mathbb{H}(\mathcal{U}_{\mathcal{K}})} \leq -\V(p_{\K},\mathcal{U}_{\mathcal{K}}) \log \frac{\V(p_{\K},\mathcal{U}_{\mathcal{K}})}{\abs{\mathcal{K}}}\\
&=
\V(p_{\K},\mathcal{U}_{\mathcal{K}}) \log \frac{2^{nR_{K}}}{\V(p_{\K},\mathcal{U}_{\mathcal{K}})}\\
&=nR_K \V(p_{\K},\mathcal{U}_{\mathcal{K}}) - \V(p_{\K},\mathcal{U}_{\mathcal{K}}) \log \V(p_{\K},\mathcal{U}_{\mathcal{K}}).
\end{align*}
}
This vanishes as long as $\V(p_{\K},\mathcal{U}_{\mathcal{K}}) \sim o\left(\frac{1}{n}\right)$, which is indeed the case.

\subsection{Strong secrecy}
\label{section_secrecy}
Using \cite[Lemma 1]{Csiszar96}, we can bound the leakage as follows:
\begin{equation} \label{leakage}
\I(\K; \mathsf{S},\Z^n,\U)=\I(\K; \mathsf{S},\hat{\Z}^n,\U) \leq d_{\av} \log \frac{\abs{\mathcal{K}}}{d_{\av}}, 
\end{equation}
where
\begin{align} 
&d_{\av}=\sum_{k \in \mathcal{K}} p_{\K}(k) \V(p_{\mathsf{S} \hat{\Z}^n \U|\K=k},p_{\mathsf{S} \hat{\Z}^n\U}) \notag \\
&=\mathbb{E}_{\hat{\Z}^n\U}\left[
\sum_{k \in \mathcal{K}} p_{\K}(k)
\mathbb{V}\left(p_{\mathsf{S}| \hat{\Z}^n \U\K=k},p_{\mathsf{S}|\hat{\Z}^n\U}\right)\right] \notag\\
& \leq \mathbb{E}_{\hat{\Z}^n\U}\left[\sum_{k \in \mathcal{K}} p_{\K}(k) \mathbb{V}\left(p_{\mathsf{S}| \hat{\Z}^n \U\K=k}, \mathcal{U}_{\mathcal{S}}\right)\right] \label{18} \\
&+ \mathbb{E}_{\hat{\Z}^n\U}\left[  \mathbb{V}\left( \mathcal{U}_{\mathcal{S}}, p_{\mathsf{S}| \hat{\Z}^n\U} \right)\right] \label{19}
\end{align}
by the triangle inequality.\\
Due to Remark \ref{bijection}, we can write
{\allowdisplaybreaks
\begin{align*}
&p_{\mathsf{S}| \hat{\Z}^n \U\K}(s|\mb{z},\mb{u},k)=\frac{p_{\mathsf{S}| \hat{\Z}^n\U\K}(s|\mb{z},\mb{u},k)}{p_{\K}(k)}\\
&=\frac{p_{\bar{\X}_Q|\hat{\Z}^n,\U}(k+s|\mb{z},\mb{u})}{p_{\K}(k)}
\\
&=
\frac{1}{p_{\K}(k)}\sum_{\lambda_3 \in \Lambda_3}p_{\X_Q|\hat{\Z}^n,\U}(k+s+\lambda_3|\mb{z},\mb{u}).
\end{align*}
}
The term (\ref{18}) can be written as
\begin{align}
& \mathbb{E}_{\hat{\Z}^n\U}\left[\sum_{k \in \mathcal{K}}\sum_{s \in \mathcal{S}}
\abs{p_{\X_Q|\hat{\Z}^n,\U}(k+s+\lambda_3)-p_{\K}(k) \frac{V(\Lambda_1)}{V(\Lambda_2)}}\right]\notag \\
&\leq \mathbb{E}_{\hat{\Z}^n\U}\left[\sum_{k \in \mathcal{K}}\sum_{s \in \mathcal{S}}
\abs{p_{\X_Q|\hat{\Z}^n,\U}(k+s+\lambda_3)-\frac{V(\Lambda_1)}{V(\Lambda_3)}}\right]
\label{18a} \\
&+ \mathbb{E}_{\hat{\Z}^n\U}\left[\sum_{k \in \mathcal{K}}\sum_{s \in \mathcal{S}}\abs{
\frac{V(\Lambda_1)}{V(\Lambda_3)}-p_{\K}(k)\frac{V(\Lambda_1)}{V(\Lambda_2)}}\right]
\label{18b}
\end{align}
by the triangle inequality.\\
Observe that 
{\allowdisplaybreaks
\begin{align*}
&p_{\X_Q|\hat{\Z}^n,\U}(\mb{x}_Q|\mb{z},\mb{u})=
\int_{\R^n} p_{\X_Q|\X^n,\U}(\mb{x}_Q|\mb{x},\mb{u})p_{\X^n|\hat{\Z}^n}(\mb{x}|\mb{z})d\mb{x}\\
&= \int_{\R^n} \frac{f_{\sigma_Q}(\mb{x}_Q-\mb{x}-\mb{u})}{f_{\sigma_Q}(\Lambda_1-\mb{x}-\mb{u})}f_{\sigma_2}(\mb{x}-\mb{z})d\mb{x}\\
&=\int_{\R^n} \frac{f_{\sigma_Q}(\mb{x}_Q-\mb{w}-\mb{z}-\mb{u})}{f_{\sigma_Q}(\Lambda_1-\mb{w}-\mb{z}-\mb{u})}f_{\sigma_2}(\mb{w})d\mb{x},
\end{align*}
}%
which is the distribution of $\nint{\W_2^n+\mb{z}+\mb{u}}_{\Lambda_1,\sigma_Q}$. \\
Therefore, the term (\ref{18a}) can be rewritten as
{\allowdisplaybreaks
\begin{align*}
& \mathbb{E}_{\hat{\Z}^n}\left[\mathbb{E}_{\U}\left[ \mathbb{V}\left(p_{\X_Q\Mod \Lambda_3|\hat{\Z}^n,\U},\mathcal{U}_{\Lambda_1/\Lambda_3}\right)\right]\right]\\
&\stackrel{(a)}{\leq} 2 \epsilon^1_{\Lambda_1}(\sigma_Q)+2\epsilon^1_{\Lambda_3}(\tilde{\sigma}_2),
\end{align*}
}%
where $\tilde{\sigma}_2^2=\sigma_2^2+\sigma_Q^2$, and (a) follows by the previous remark and Corollary \ref{corollary_randomized_rounding}.
This vanishes as $o\left(\frac{1}{n}\right)$ assuming the conditions (\ref{condition_Lambda1}) and (\ref{condition_Lambda3}).\\
The term (\ref{18b}) simplifies to
$$\sum_{k \in \mathcal{K}} \abs{\frac{V(\Lambda_2)}{V(\Lambda_3)}-p_{\K}(k)}=\mathbb{V}(\mathcal{U}_{\mathcal{K}},p_{\K})=o\left(\frac{1}{n}\right) \to 0$$
as already shown in Section \ref{uniformity_section}.\\
Observe that
\begin{align*}
&p_{\S\hat{\Z}^n\U}(s,\mb{z},\mb{u})=\sum_{k'\in \mathcal{K}} p_{\S\K\hat{\Z}^n\U}(s,k',\mb{z},\mb{u})\\
&=
\sum_{k'\in \mathcal{K}}\frac{p_{\hat{\Z}^n}(\mb{z})}{V(\Lambda_1)} p_{\bar{\X}_Q|\hat{\Z}^n\U}(s+k'|\mb{z},\mb{u})\\
&=
\sum_{k'\in \mathcal{K}}\frac{p_{\hat{\Z}^n}(\mb{z})}{V(\Lambda_1)} \sum_{\lambda_3 \in \Lambda_3} p_{\X_Q|\hat{\Z}^n\U}(s+k'+\lambda_3|\mb{z},\mb{u})
\end{align*}
We now come back to the expression (\ref{19}), which is equal to
{\allowdisplaybreaks
 \begin{align*}
&\mathbb{E}_{\hat{\Z}^n\U}\!\left[\sum_{s \in \mathcal{S}}  \abs{
\frac{V(\Lambda_1)}{V(\Lambda_2)}\!-\!\smash{\sum_{k'\in \mathcal{K}} \sum_{\lambda_3 \in \Lambda_3}}\! p_{\X_Q|\hat{\Z}^n,\U}(s\!+\!k'\!+\!\lambda_3)}\right] \\
&\!\leq\! \mathbb{E}_{\hat{\Z}^n\U}\!\left[\sum_{k'\in \mathcal{K}}\sum_{s \in \mathcal{S}}  \abs{
\frac{V(\Lambda_1)}{V(\Lambda_3)}\!-\!\! \smash{\sum_{\lambda_3 \in \Lambda_3}}\! p_{\X_Q|\hat{\Z}^n,\U}(s\!+\!k'\!+\!\lambda_3)}\right] \\
&\!=\!\mathbb{E}_{\hat{\Z}^n}\left[
\mathbb{E}_{\U}\left[\V\left(\mathcal{U}_{\Lambda_1/\Lambda_3},p_{\X_Q|\hat{\Z}^n,\U}\Mod\Lambda_3\right)\right]\right]
\\
&\leq 2\epsilon^1_{\Lambda_1}(\sigma_Q)+2\epsilon^1_{\Lambda_3}(\tilde{\sigma}_2)
\end{align*} 
}
by Corollary \ref{corollary_randomized_rounding}. 
This again vanishes as $o\left(\frac{1}{n}\right)$ under conditions (\ref{condition_Lambda1}) and (\ref{condition_Lambda3}).\\
In conclusion, $d_{\av} \sim o\left(\frac{1}{n}\right)$ and thus from (\ref{leakage}), we find that the leakage vanishes asymptotically as $n \to \infty$. 

\begin{rem}
Although in Section \ref{uniformity_section} we only showed that the key is close to uniform on average over the dither $\U$, using the results in this section we see that
\begin{align*}
&\H(\mathcal{U}_{\mathcal{K}})-\H(\K|\U)=\H(\mathcal{U}_{\mathcal{K}})-\H(\K)+\I(\K;\U)\\
&\leq \H(\mathcal{U}_{\mathcal{K}})-\H(\K)+\I(\K;\mathsf{S},\Z^n,\U)  \to 0.
\end{align*}
\end{rem}

\subsection{Achievable strong secrecy rate and optimal trade-off}
\label{section_trade_off}
Recall that in the previous sections we have imposed the conditions (\ref{condition_Lambda1}), (\ref{condition_Lambda2}) and (\ref{condition_Lambda3}) on the volumes of $\Lambda_1$, $\Lambda_2$ and $\Lambda_3$ respectively, i.e.
$$\frac{V(\Lambda_1)^{2/n}}{\sigma_Q^2}< 2 \pi e, \quad \frac{V(\Lambda_2)^{2/n}}{\tilde{\sigma}_1^2} > 2\pi e, \quad \frac{V(\Lambda_3)^{2/n}}{\tilde{\sigma}_2^2} < 2 \pi e.$$
Therefore, the achievable secret key rate is upper bounded by
\begin{equation} \label{bound_R_K}
R_K =\frac{1}{n} \log\frac{V(\Lambda_3)}{V(\Lambda_2)} < \frac{1}{2} \log \frac{\tilde{\sigma}_2^2}{\tilde{\sigma}_1^2}=\frac{1}{2} \log \frac{\sigma_2^2+\sigma_Q^2}{\sigma_1^2+\sigma_Q^2}
\end{equation}
As $\sigma_Q \to 0$, 
$$R_K \to \frac{1}{2}\log \frac{\sigma_2^2}{\sigma_1^2},$$
which is the optimal secret key rate. This improves upon our previous work \cite{LLB13} in which the achievable secrecy rate had a $1/2$ nat gap compared to the optimal.

\begin{rem}
The optimal scaling of the lattice $\Lambda_3$ requires the noise variance $\sigma_2$ to be known by Alice; if only a lower bound for $\sigma_2$ is available, positive secret key rates can still be attained.
\end{rem}

The public communication rate is lower bounded by
$$R_P=\frac{1}{n} \log \frac{V(\Lambda_2)}{V(\Lambda_1)} > \frac{1}{2} \log \frac{\sigma_1^2+\sigma_Q^2}{\sigma_Q^2}.$$ 
Equivalently, we have $\sigma_Q^2>\frac{\sigma_1^2}{e^{2R_P}-1}$. Replacing this expression in the bound (\ref{bound_R_K}) for $R_K$, and observing that (\ref{bound_R_K}) is a decreasing function of $\sigma_Q^2$, we find
$$R_K < \frac{1}{2} \log \left(e^{-2R_P}+\frac{\sigma_2^2}{\sigma_1^2}(1-e^{-2R_P})\right).$$
which corresponds to the optimal public rate / secret key rate trade-off (\ref{trade_off_definition}).

\section{Conclusions and perspectives} \label{conclusion}

To conclude, we have 
proposed a new lattice-based technique 
to extract a secret key from correlated Gaussian sources against an eavesdropper. Using $L^1$ distance and KL divergence, we have proved the existence of lattices with a vanishing flatness factor for all VNRs up to $2\pi e$. This improves upon the previous result for VNRs up to $2\pi$, based on $L^{\infty}$ distance. Together with dithering and randomized rounding, it has enabled us to achieve the optimal trade-off with one-way public communication.
In the same way, it is possible to remove the $\frac{1}{2}$-nat gap to the secrecy capacity of wiretap channels \cite{LLBS} associated to the use of the $L^{\infty}$ flatness factor \cite[p. 1656]{LiuLing18}.

An immediate step for future work is to turn the existence result of this paper into a practical scheme. 
There are avenues for replacing random nested lattices for Wyner-Ziv coding with lower-complexity techniques, such as superposition coding or residual quantization \cite{bennatan2006superposition,wei2016residual}. However such techniques do not address privacy amplification.  
In order to implement the approach proposed in this paper based on the notion of flatness factor of a lattice, a promising option is to
instantiate the lattices using polar codes (aka polar lattices), which have been shown to be good for quantization, channel coding \cite{LiuYanLingWu2019} and secrecy.
A polar lattice has been constructed in \cite{LiuLing18} to achieve the secrecy capacity of Gaussian wiretap channels. It can be shown that the secrecy-good lattice in \cite{LiuLing18} enjoys a vanishing $L^{1}$ flatness factor. Since the encoding and decoding complexity of a polar lattice is quasi-linear in blocklength $n$, it is an excellent candidate to build a practical scheme for secret key generation. It is also possible to implement the randomized rounding algorithm over a polar lattice. We leave such implementation issues to future work.

Another problem is to see if it is possible to modify the design of this paper to yield a fuzzy extractor, which would require redesigning a lattice with respect to other entropy measures. Other open problems include identifying whether is is possible to remove dithering and/or randomized quantization, characterizing the second-order asymptotics and the extension of the proposed key-agreement protocol to multi-terminal systems. Furthermore, the reconciliation technique based on Wyner-Ziv coding may be extended to key-encapsulation mechanisms  (KEM) in lattice-based cryptography, due to the similarity between KEM and secret key agreement. Finally, it is interesting to explore the applications of $L^1$ and KL smoothing parameters in other cryptographic and mathematical problems \cite{Chung2013,Dadush_Regev_2016}.

\section*{Acknowledgments}
The second author is grateful to Antonio Campello, Daniel Dadush and Ling Liu for helpful discussions.
The authors would like to thank the two anonymous reviewers for their detailed comments and suggestions which helped improve the paper. 

\appendices

	\section{Resolvability codes} \label{appendix_resolvability}
	In this section we review some results from \cite{Hayashi_Matsumoto2016} about resolvability codes for regular channels, which are needed for the proof of Theorem \ref{VNR_L1}. 
	
	First, we need some preliminary definitions. In the following, we assume $\mathcal{X}$ is a finite abelian group and $\mathcal{Y}$ is a measurable space. Given a channel $W: \mathcal{X} \to \mathcal{Y}$, we use the notation $W_x(y)=W(y|x)$ for $x \in \mathcal{X}, y \in \mathcal{Y}$.
	
	\begin{deft}[Rényi Entropy]
	Given a discrete distribution $p_{\mathsf{A}}$ on $\mathcal{A}$ and $\rho \geq 0$, we define 
	$$H_{1+\rho}(\mathsf{A})=-\frac{1}{\rho} \log \sum_{a \in \mathcal{A}} p_{\mathsf{A}}(a)^{1+\rho}.$$ 
	\end{deft}
	
	\begin{deft} \label{Hayashi_function}
	Given a channel $W: \mathcal{X} \to \mathcal{Y}$ and a probability distribution $p_{\X}$ on $\mathcal{X}$, we define $\forall \rho \geq 0$
	$$\psi(\rho| W, p_{\X})=\log \sum_{x \in \mathcal{X}} p_{\X}(x) \int_{\mathcal{Y}} W_x(y)^{1+\rho} (W \circ p_X)(y)^{-\rho} dy.$$
	\end{deft}
	
	This function has the following properties:
	\begin{align}
	& \psi(0|W,p_{\X})=0 \label{eq_psi_0}, \\
	&\psi(\rho|W^n, p_{\X}^{\otimes n})=n\psi(\rho|W,p_{\X}) \label{eq_psi_1}, \\
	& \lim_{\rho \to 0} \frac{\psi(\rho|W,p_{\X})}{\rho}=\I(\X;\Y). \label{eq_psi_2}
	\end{align}
	
	We also compute the second derivative in $0$ which will be needed in the next section.
			\begin{lem} \label{second_derivative}
		\begin{align*}
		&\psi''(0)=\sum_{x \in \mathcal{X}} p_{\X}(x)\int_{\mathcal{Y}}W_x(y)\left(\log \frac{W_x(y)}{(W \circ p_{\X}(y))}\right)^2dy\\
		&-\left(\sum_{x \in \mathcal{X}} p_{\X}(x) \int_{\mathcal{Y}}W_x(y) \log\frac{W_x(y)}{(W \circ p_{\X})(y)}dy\right)^2.
		\end{align*}
		\end{lem}
The proof of Lemma \ref{second_derivative} can be found in Appendix \ref{proof_second_derivative}.

	\begin{deft}[Regular channel] \label{deft_regular}
		 The channel $W: \mathcal{X} \to \mathcal{Y}$ is called \emph{regular} if $\mathcal{X}$ acts on $\mathcal{Y}$ by permutations $\{\pi_x\}_{x \in \mathcal{X}}$ such that $\pi_x(\pi_x'(y))=\pi_{x+x'}(y)$ $\forall x,x' \in \mathcal{X}$, and there exists a probability density $p_{\Y}$ on $\mathcal{Y}$ such that $W_x(y)=p_{\Y}(\pi_x(y))$ $\forall x \in \mathcal{X}$, $\forall y \in \mathcal{Y}$.   \end{deft}
		 
In particular, a regular channel is symmetric \cite{Forney00,Delsarte_Piret} in the sense of Gallager \cite{Gallager}, and its capacity is achieved by the uniform distribution. 

The following theorem was stated for discrete memoryless channels \cite[Corollary 18]{Hayashi_Matsumoto2016} but can be extended to continuous outputs \cite[Appendix D]{Hayashi_Matsumoto2016} as follows:
		
		\begin{theorem} \label{theorem_resolvability}
		Let $\mathcal{M}$ and $\mathcal{X}$ be a finite-dimensional vector spaces over $\mathbb{F}_p$ and $\mathcal{Y}$ a measurable space.
		Consider a uniform random variable $\mathsf{F}$ taking values over the set of  
		linear mappings $f:\mathcal{M} \to \mathcal{X}$ and a distribution $p_{\mathsf{M}}$ on $\mathcal{M}$. 
		If $W : \mathcal{X} \to \mathcal{Y}$ is regular, then $\forall \rho \in(0,1]$,
		$$\mathbb{E}_{\mathsf{F}}\left[e^{\rho \mathbb{D}(W \circ \mathsf{F} \circ p_{\mathsf{M}} || W \circ \mathcal{U}_{\mathcal{X}})}\right] \leq 1+ e^{-\rho H_{1+\rho}(\mathsf{M})}e^{\psi(\rho|W,\mathcal{U}_{\mathcal{X}})}.$$ 
		\end{theorem}
		
		Theorem \ref{theorem_resolvability} is a one-shot result, but we can apply it to $n$ uses of an i.i.d. channel to get the following. 
		
\begin{corol} \label{corol_resolvability}
Let $\mathcal{X}$ be a finite-dimensional vector space over $\mathbb{F}_p$ and $\mathcal{Y}$ a measurable space, and 
 $W: \mathcal{X} \to \mathcal{Y}$ a regular channel. Let $R>\I(\mathsf{X};\mathsf{Y})$, where $\mathsf{X} \sim \mathcal{U}_{\mathcal{X}}$ and $\mathsf{Y} \sim W \circ \mathcal{U}_{\mathcal{X}}$. 
Consider $\mathsf{C}_n \subset \mathcal{X}^n$ chosen uniformly at random in the set of $(n,k)$ linear codes in $\mathcal{X}^n$, where $k=\frac{\ceil{nR}}{\log p}$. Denote by $\mathcal{U}_{\mathsf{C}_n}$ the uniform distribution over the codewords in $\mathsf{C}_n$. Then  
$$\mathbb{E}_{\mathsf{C}_n}[\mathbb{D}(W^n \circ \mathcal{U}_{\mathsf{C}_n}||W^n \circ \mathcal{U}_{\mathcal{X}}^{\otimes n})] \to 0$$
exponentially fast as $n \to \infty$. 
\end{corol}
\begin{IEEEproof}
Note that  $W^n : \mathcal{X}^n \to \mathcal{Y}^n$ is still a regular channel with respect to the set of permutations $\{\pi_{\mb{x}}\}_{\mb{x} \in \mathcal{X}^n}$, where we define
$\pi_{\mb{x}}(y_1,\ldots,y_n)=(\pi_{x_1}(y_1),\ldots,\pi_{x_n}(y_n))$ for $\mb{x}=(x_1,\ldots,x_n)$.\\
Applying Theorem \ref{theorem_resolvability} to this channel, and taking $\mathcal{M}=\mathbb{F}_p^{k}$ with $k=\frac{\ceil{nR}}{\log p}$ and $p_{\mathsf{M}}=\mathcal{U}_{\mathcal{M}}$, for  $\mathsf{F}_n$ representing a uniform random linear encoder $f_n :\mathcal{M} \to \mathcal{X}^n$ we have
\begin{align*}
&\mathbb{E}_{\mathsf{F}_n}\left[e^{\rho \mathbb{D}(W^n \circ \mathsf{F}_n \circ \mathcal{U}_{\mathcal{M}} || W^n \circ \mathcal{U}_{\mathcal{X}}^{\otimes n})}\right] \\
&\leq 1+ e^{-\rho H_{1+\rho}(\mathsf{M})}e^{\psi(\rho|W^n,\mathcal{U}_{\mathcal{X}}^{\otimes n})}. 
\end{align*}
By Jensen's inequality,
\begin{align*}
&\mathbb{E}_{\mathsf{F}_n}[\mathbb{D}(W^n \circ \mathsf{F}_n \circ \mathcal{U}_{\mathcal{M}}||W^n \circ \mathcal{U}_{\mathcal{X}}^{\otimes n})] \\
&\leq \frac{1}{\rho} \log \left(1+e^{-\rho H_{1+\rho}(\mathsf{M})} e^{\psi(\rho|W^n, \mathcal{U}_{\mathcal{X}}^{\otimes n})}\right)\\
&\leq \frac{1}{\rho} e^{-\rho H_{1+\rho}(\mathsf{M})+\psi(\rho|W^n, \mathcal{U}_{\mathcal{X}}^{\otimes n})}.
\end{align*}
Note that $H_{1+\rho}(\mathsf{M})=nR$ since $\M$ is uniform. Using (\ref{eq_psi_1}), we find that $\forall \rho \in (0,1]$,
\begin{align} 
&\mathbb{E}_{\mathsf{F}_n}[\mathbb{D}(W^n \circ \mathsf{F}_n \circ \mathcal{U}_{\mathcal{M}}||W^n \circ \mathcal{U}_{\mathcal{X}}^{\otimes n})] \notag \\
&\leq 
\frac{1}{\rho} e^{-n(\rho R-\psi(\rho|W, \mathcal{U}_{\mathcal{X}}))}.\label{E_Fn}
\end{align}
From (\ref{eq_psi_0}) and (\ref{eq_psi_2}), we have $\psi(\rho | W, p_{\X})= \rho \I(\X;\Y) + \eta(\rho)$, where $\lim_{\rho \to 0} \frac{\eta(\rho)}{\rho}=0$. Given $R> \I(\X;\Y)$, $\exists \bar{\rho}$ sufficiently small such that $\delta=R-\I(\X;\Y)-\frac{\eta(\bar{\rho})}{\bar{\rho}}>0$. Therefore
\begin{equation} \label{rho_bar}
\mathbb{E}_{\mathsf{F}_n}[\mathbb{D}(W^n \circ \mathsf{F}_n \circ \mathcal{U}_{\mathcal{M}}||W^n \circ \mathcal{U}_{\mathcal{X}}^{\otimes n})] \leq \frac{1}{\bar{\rho}} e^{-n \bar{\rho} \delta} \to 0
\end{equation}
as $n \to \infty$. 
The conclusion follows by noting that $\mathsf{F}_n \circ \mathcal{U}_{\mathcal{M}}=\mathcal{U}_{\mathsf{C}_n}$.
\end{IEEEproof}

\section{Modulo lattice channels and the KL flatness factor}

\label{Appendix_KL}
In this section, we review some properties of modulo lattice channels and introduce another notion of flatness factor based on the KL divergence, which will be used in the proof of Theorem 
\ref{VNR_L1}.

\subsection{The mod-$\Lambda$ channel and the mod-$\Lambda/\Lambda'$ channel} \label{subsection_Forney}

Following Forney et al. \cite{Forney00}, given a fundamental region $\mathcal{R}(\Lambda)$ of a lattice $\Lambda$ we can define the mod-$\Lambda$ channel with input $\X^n \in \mathcal{R}(\Lambda)$ and output
$$\Y^n=[\X^n + \mathsf{W}^n] \Mod \mathcal{R}(\Lambda),$$
where $\mathsf{W}^n$ is a noise vector. When $\mathsf{W}^n$ is i.i.d. Gaussian with variance $\sigma^2$, this channel has capacity 
$$C(\Lambda, \sigma^2)=\log V(\Lambda)-h(f_{\sigma,\Lambda}).$$

In the above expression, with slight abuse of notation we denote by $h(f_{\sigma,\Lambda})$ the differential entropy of $f_{\sigma,\mathcal{R}(\Lambda)}$, which does not depend on the choice of the region $\mathcal{R}(\Lambda)$.

The following result \cite[Lemma 1]{LiuYanLingWu2019} relates the $L^{\infty}$ flatness factor to the capacity of the mod-$\Lambda$ channel.

\begin{lem} \label{Lemma1_LYLW}
The capacity $C(\Lambda,\sigma^2)$ of the mod-$\Lambda$ channel is bounded by
$C(\Lambda,\sigma^2)\leq \log(1+\epsilon_{\Lambda}(\sigma))\leq \epsilon_{\Lambda}(\sigma).$
\end{lem}

Given two nested lattices $\Lambda'\subset \Lambda$ and a fundamental region $\mathcal{R}(\Lambda')$, we can define the mod-$\Lambda/\Lambda'$ channel with discrete input $\X^n \in \Lambda \cap \mathcal{R}(\Lambda')$ and output
$$\Y^n=[\X^n + \mathsf{W}^n] \Mod \mathcal{R}(\Lambda').$$
It was shown in \cite{Forney00} that this channel has capacity
$$C(\Lambda/\Lambda', \sigma^2)=\log \abs{\Lambda / \Lambda'}+h(f_{\sigma,\Lambda})-h(f_{\sigma,\Lambda'}).$$
In particular, the following relation holds:
\begin{equation} \label{Forney_equation}
C(\Lambda/\Lambda',\sigma^2)=C(\Lambda',\sigma^2)-C(\Lambda,\sigma^2).
\end{equation}

\begin{lem} \label{capacity_lemma}
For any $\sigma>0$,
$$C(\Lambda/\Lambda',\sigma^2)=\mathbb{D}\left(f_{\sigma,\mathcal{R}(\Lambda')}\Big\Vert \frac{1}{\abs{\Lambda/\Lambda'}}{f_{\sigma,\Lambda}}_{|\mathcal{R}(\Lambda')}  \right).$$
\end{lem}

\begin{IEEEproof}
By definition,
\begin{align*}
&\mathbb{D}\left(f_{\sigma,\mathcal{R}(\Lambda')}\Big\Vert \frac{1}{\abs{\Lambda/\Lambda'}}{f_{\sigma,\Lambda}}_{|\mathcal{R}(\Lambda')}  \right)\\
&=\int_{\mathcal{R}(\Lambda')} f_{\sigma,\Lambda'}(\mb{y}) \log \frac{f_{\sigma,\Lambda'}(\mb{y}) \abs{\Lambda/\Lambda'}}{f_{\sigma,\Lambda}(\mb{y})} d\mb{y}\\
&=-h(f_{\sigma,\Lambda'}) + \int_{\mathcal{R}(\Lambda')} f_{\sigma,\Lambda'}(\mb{y}) \log \frac{\abs{\Lambda/\Lambda'}}{f_{\sigma,\Lambda}(\mb{y})} d\mb{y}\\
&=-h(f_{\sigma,\Lambda'}) + \log \abs{\Lambda/\Lambda'} - \int_{\mathcal{R}(\Lambda')} f_{\sigma,\Lambda'}(\mb{y}) \log f_{\sigma,\Lambda}(\mb{y}) d\mb{y}.  
\end{align*}
The conclusion follows by observing that
{\allowdisplaybreaks
\begin{align}
&- \int_{\mathcal{R}(\Lambda')} f_{\sigma,\Lambda'}(\mb{y}) \log f_{\sigma,\Lambda}(\mb{y}) d\mb{y} \notag\\
&=-\sum_{\lambda \in \Lambda / \Lambda'} \int_{\mathcal{R}(\Lambda)+\lambda} f_{\sigma,\Lambda'}(\mb{y}) \log f_{\sigma,\Lambda}(\mb{y}) d\mb{y} \notag\\
=&-\sum_{\lambda \in \Lambda / \Lambda'} \int_{\mathcal{R}(\Lambda)} f_{\sigma,\Lambda'}(\mb{y}-\lambda)\log f_{\sigma,\Lambda}(\mb{y}) d\mb{y} \notag\\
&=-\int_{\mathcal{R}(\Lambda)} f_{\sigma,\Lambda}(\mb{y})\log f_{\sigma,\Lambda}(\mb{y}) d\mb{y}=h(f_{\sigma,\Lambda}). \tag*{\IEEEQED}
\end{align}
}
\let\IEEEQED\relax%
\end{IEEEproof}

\subsection{The KL flatness factor}
We can now introduce a notion of flatness factor based on KL divergence.

\begin{deft}
	Given a lattice $\Lambda$, a fundamental region $\mathcal{R}(\Lambda)$ and $\sigma>0$, we define the \emph{KL flatness factor} as follows:
	\begin{equation} \label{KL_flatness_factor}
	\epsilon^{\KL}_{\Lambda}(\sigma)=\mathbb{D}(f_{\sigma,\mathcal{R}(\Lambda)}||\mathcal{U}_{\mathcal{R}(\Lambda)}).
\end{equation}
\end{deft}

Note that as before, the definition does not depend on the choice of the fundamental region. 

\begin{rem} \label{Pinsker}
By Pinsker's inequality, $\forall \sigma >0$,
$$\epsilon_{\Lambda}^{1}(\sigma) \leq \sqrt{2 \epsilon_{\Lambda}^{\KL}(\sigma)}.$$
\end{rem}

\begin{rem}[Relation to the capacity of the mod-$\Lambda$ channel] \label{remark_capacity_mod_Lambda}
	Note that \cite[p.1656]{LiuLing18}
	$$\mathbb{D}(f_{\sigma,\mathcal{R}(\Lambda)}||\mathcal{U}_{\mathcal{R}(\Lambda)})=\log V(\Lambda)-h(f_{\sigma,\Lambda})=C(\Lambda,\sigma^2).$$
	\end{rem}
By shift-invariance of the differential entropy, the KL flatness factor is also shift-invariant, i.e. $$\epsilon_{\Lambda}^{\KL}(\sigma)=\mathbb{D}({f_{\sigma,\Lambda, \mb{c}}}_{|\mathcal{R}(\Lambda)}||\mathcal{U}_{\mathcal{R}(\Lambda)})$$ for all $\mb{c} \in \R^n$. 

Thanks to Remark \ref{remark_capacity_mod_Lambda}, we are able to prove that the KL flatness factor is monotonic: 
\begin{lem} \label{KL_monotonic}
For any lattice $\Lambda$, $\forall \sigma'>\sigma$,
	$\epsilon^{\KL}_{\Lambda}(\sigma') \leq \epsilon^{\KL}_{\Lambda}(\sigma).$
	\end{lem}
	\begin{IEEEproof}
With the same notation as in the proof of Lemma 
\ref{L1_monotonic}, from the data processing inequality for the KL divergence \cite[Lemma 3.11]{Csiszar_Korner} we have
\begin{align}
&\epsilon^{\KL}_{\Lambda}\left(\sqrt{\sigma^2+\sigma_0^2}\right)=
\mathbb{D}\left(f_{\sqrt{\sigma^2+\sigma_0^2},\mathcal{R}(\Lambda)}||\mathcal{U}_{\mathcal{R}(\Lambda)}\right)\notag \\
&=
\D(\Y^n || \mathsf{U}^n) \leq \D(\X^n || \mathsf{U}^n)= 
 \mathbb{D}(f_{\sigma,\mathcal{R}(\Lambda)}||\mathcal{U}_{\mathcal{R}(\Lambda)})\\
 &=\epsilon^{\KL}_{\Lambda}(\sigma).\tag*{\IEEEQED}
 \end{align}
\let\IEEEQED\relax%
\end{IEEEproof}

Similarly to Definition \ref{L1_secrecy_good}, we can introduce a notion of secrecy goodness based on the KL flatness factor. 
\begin{deft} \label{KL_secrecy_good}
A sequence of lattices $\{\Lambda^{(n)}\}$ is \emph{KL secrecy-good} if $\epsilon^{\KL}_{\Lambda^{(n)}}(\sigma) =o\left(\frac{1}{n^c}\right).$
\end{deft}
\begin{rem} \label{Pinsker2}
By Remark \ref{Pinsker}, a sequence of KL secrecy-good lattices is also $L^1$ secrecy-good. 
\end{rem}%

One can show that under the assumption of a small KL flatness factor, the modulo lattice operation 
allows to extract the \emph{intrinsic randomness} of the additive Gaussian channel (in the sense of \cite{Bloch_Intrinsic_Randomness}). The interested reader can find more details in the preprint version of this work \cite{secret_key_preprint}.

\section{Proof of Theorem \ref{VNR_L1}} \label{proof_VNR_L1}

In order to prove Theorem \ref{VNR_L1}, we will actually show a stronger result.	
\begin{prop} \label{VNR_KL}
If $\gamma_{\Lambda}(\sigma)<2 \pi e$ is fixed, then there exists a sequence $\{\Lambda^{(n)}\}$ of lattices which are KL secrecy-good.
\end{prop}
Theorem \ref{VNR_L1} then follows from Proposition \ref{VNR_KL} by Remark \ref{Pinsker2}.

Before proceeding with the proof, we summarize the main idea here. We use the standard Construction A to find the sought-after lattice $\Lambda$, by choosing a coarse lattice $\Lambda_c=\alpha p \mathbb{Z}$, a fine lattice $\Lambda_f=\alpha \mathbb{Z}$, an $(n,k)$ linear code $\mathcal{C}$ over $\mathbb{F}_p$, and 
$\Lambda_c^n \subseteq \Lambda=\alpha(p\mathbb{Z}^n + \mathcal{C}) \subseteq \Lambda_f^n$. 
Using the chain rule (\ref{Forney_equation}), we have 
		$$\mathbb{D}(f_{\mathcal{R}(\Lambda),\sigma}|| \mathcal{U}_{\mathcal{R}(\Lambda)})=C(\Lambda,\sigma^2)=C(\Lambda_f^n,\sigma^2)+C(\Lambda_f^n/\Lambda,\sigma^2).$$
Now, using a sufficiently fine lattice $\Lambda_f$, we can easily make $C(\Lambda_f^n,\sigma^2)\to 0$ thanks to the flatness phenomenon (cf. Lemma~\ref{Lemma1_LYLW}). The non-trivial part of the proof is to exhibit a lattice $\Lambda$  such that $C(\Lambda_f^n/\Lambda,\sigma^2)\to 0$ as well. 
It turns out that if the linear code $\mathcal{C}$ is a \emph{resolvability code} for the mod-$\Lambda_f/\Lambda_c$ channel $W$, i.e. if the output of the code is close to the output of uniform input, then $\mathcal{C}$ provides the desired solution. In fact, we show that
$$\mathbb{D}(W^n \circ \mathcal{U}_{\mathcal{C}} || W^n \circ \mathcal{U}_{(\Lambda_f/\Lambda_c)^n})=C(\Lambda_f^n/\Lambda,\sigma^2),$$
which tends to $0$ if $\mathcal{C}$ is a resolvability code. The existence of such linear resolvability codes follows from the results of \cite{Hayashi_Matsumoto2016} (see Appendix \ref{appendix_resolvability}).
However, making the above argument rigorous involve certain technicalities, as seen in the following.

	\begin{IEEEproof}[Proof of Proposition \ref{VNR_KL}]
	
		For a given dimension $n$, we will construct $\Lambda$ as a scaled mod-$p$ lattice \cite{Loeliger} of the form $\Lambda=\alpha(p \mathbb{Z}^n+\mathcal{C}_n)$, where $\mathcal{C}_n$ is an $(n,k)$-linear code over $\mathbb{F}_p$. 
		
		We will consider the asymptotic behavior as $n \to \infty$, $\alpha \to 0, p \to \infty$ while satisfying the volume condition $\alpha^n p^{n-k}=V(\Lambda)=(\gamma \sigma^2)^{n/2}$. Here, $\gamma$ is the volume-to-noise ratio, which is assumed to be fixed.
		
		By construction, $\Lambda_c^{n}\subset \Lambda \subset \Lambda_f^{n}$, where $\Lambda_c=\alpha p \mathbb{Z}$ and $\Lambda_f=\alpha \mathbb{Z}$ are one-dimensional lattices.

	From Remark \ref{remark_capacity_mod_Lambda} and the relation (\ref{Forney_equation}), we have 
		$$\mathbb{D}(f_{\sigma,\mathcal{R}(\Lambda)}|| \mathcal{U}_{\mathcal{R}(\Lambda)})=C(\Lambda,\sigma^2)=C(\Lambda_f^n,\sigma^2)+C(\Lambda_f^n/\Lambda,\sigma^2).$$
		We want to show that both terms in the sum tend to zero when $n \to \infty$. 
		
			First, we will show that $C(\Lambda_f^n,\sigma^2)=C((\alpha \mathbb{Z})^n, \sigma^2) \to 0$ 
			if $\alpha =o\left(\frac{1}{n^{c}}\right)$ for some $c>0$.
			We follow the same approach as in \cite[Appendix A]{LiuYanLingWu2019}. From Lemma \ref{Lemma1_LYLW} we have that $C(\Lambda_f^n,\sigma^2)\leq \epsilon_{\Lambda_f^n}(\sigma)$. Furthermore, it was shown in \cite[Lemma 3]{Ling_Belfiore} that
			\begin{equation} \label{flatness_factor_Zn}
			\epsilon_{\Lambda_f^n}(\sigma)=(1+\epsilon_{\Lambda_f}(\sigma))^n-1.
			\end{equation}
			Finally, one can show that \cite[Appendix A]{LiuYanLingWu2019} 
			\begin{equation} \label{flatness_factor_bound}
			\epsilon_{\Lambda_f}(\sigma)=\epsilon_{\alpha \mathbb{Z}}(\sigma) \leq 4 e^{-\frac{2 \pi ^2 \sigma^2}{\alpha^2}}.
			\end{equation} 
			Then
			{\allowdisplaybreaks
			\begin{align*}
			&\epsilon_{\Lambda_f^n}(\sigma) \leq \left(1+4 e^{-\frac{2 \pi ^2 \sigma^2}{\alpha^2}}\right)^n-1\\ 
			&\leq 4ne^{-\frac{2 \pi ^2 \sigma^2}{\alpha^2}}+o(e^{-\frac{2 \pi ^2 \sigma^2}{\alpha^2}}) \to 0.
			\end{align*}
			}
			since $(1+x)^n=1+nx+o(x)$ when $x \to 0$. 
			Next, we want to show that there exists a sequence of lattices $\Lambda$ of the form $\alpha(p\mathbb{Z}^n+\mathcal{C}_n)$ such that $C(\Lambda_f^n/\Lambda,\sigma^2) \to 0$ as $n \to \infty$. \\
			Consider the mod-($\Lambda_f/\Lambda_c$) channel $W: \Lambda_f \cap \mathcal{R}(\Lambda_c) \to \mathcal{R}(\Lambda_c)$. This channel is regular (see Definition \ref{deft_regular} in Appendix \ref{appendix_resolvability}) with respect to the set of permutations $\pi_x(y)=[y-x] \mod \Lambda_c$ for $x \in \mathcal{X}= \Lambda_f \cap \mathcal{R}(\Lambda_c)$, $y \in \mathcal{R}(\Lambda_c)$. In fact,
			\begin{align*}
			&W_x(y)=W(y|x)=f_{\sigma,\Lambda_c}(y-x)\\
			&=f_{\sigma,\Lambda_c}([y-x] \Mod \Lambda_c)=f_{\sigma,\Lambda_c}(\pi_x(y)). 
			\end{align*}
			Being regular, the mod $\Lambda_f/\Lambda_c$ channel is symmetric and the uniform distribution over $\mathcal{X}$ achieves capacity (see Appendix \ref{appendix_resolvability}).
			Moreover, $\Lambda_f/\Lambda_c \cong \mathbb{F}_p$ as abelian groups.
			We consider the required rate condition in Corollary 1:	
			\begin{align} 
			&R=\frac{1}{n} \log \abs{\mathcal{C}_n}=\frac{1}{n} \log \abs{\Lambda / \Lambda_c^n}		
			=\frac{1}{n} \log \frac{\alpha^n p^n}{V(\Lambda)} \notag \\
			&> \I(\X;\Y)=C(\Lambda_f/\Lambda_c,\sigma^2). \label{rate_condition}
			\end{align}				
			 We have
			\begin{align*}
			&C(\Lambda_f/\Lambda_c,\sigma^2)=\log \abs{\Lambda_f/\Lambda_c}+h(f_{\sigma,\Lambda_f})-h(f_{\sigma,\Lambda_c})\\
			&=\log p+h(f_{\sigma,\Lambda_f})-h(f_{\sigma,\Lambda_c})\\
			&=\log p + \log \alpha -C(\Lambda_f,\sigma^2) - h(f_{\sigma,\Lambda_c}).
			\end{align*}
			Therefore, the condition (\ref{rate_condition}) is equivalent to 
			$$ \frac{1}{n}\log V(\Lambda) < h(f_{\sigma,\Lambda_c})+C(\Lambda_f,\sigma^2).$$
			In the asymptotic limit for $\alpha \to 0$, $p\to \infty$ while  keeping $\alpha^n p^{n-k}=V(\Lambda)=(\gamma \sigma^2)^{n/2}$, we have $C(\Lambda_f,\sigma^2) \to 0$. Moreover,  $\alpha p \to \infty$, and so
			$h(\Lambda_c,\sigma^2) \to \frac{1}{2} \log 2\pi e \sigma^2$. So asymptotically, the rate condition is satisfied when
			\begin{align} \label{VNR_condition} 
			\frac{V(\Lambda)^{2/n}}{2 \pi e \sigma^2}<1.
			\end{align}		
			In this case we have 
			{\allowdisplaybreaks
			\begin{align} 
			& R-\mathbb{I}(\X;\Y) \notag\\ &=-\frac{1}{n} \log V(\Lambda)+C(\Lambda_f,\sigma^2)-h(f_{\sigma,\Lambda_c}) 
			\to \delta_0\notag \\ 
			&= \frac{1}{2}\log \frac{2\pi e \sigma^2}{V(\Lambda)^{2/n}}=\frac{1}{2}\log \frac{2\pi e}{\gamma_{\Lambda}(\sigma)}>0 \label{delta_0}
			\end{align}
			}
			as $n \to \infty$. 
			\begin{rem}
			\label{remark_convergence}
			Note that we cannot directly apply  Corollary \ref{corol_resolvability} in Appendix \ref{appendix_resolvability} to this setting, since the definition of the channel $W$ depends on $\alpha$ and $p$ which are not fixed but are a function of $n$. However, we will show that the proof of the Corollary can be extended to this channel since the convergence in (\ref{rho_bar}) is uniform. 
			\end{rem}

		\begin{IEEEproof}[Proof of Remark \ref{remark_convergence}]
		Let $\mathsf{X}$ be a uniformly distributed variable on $\Lambda_f \cap \mathcal{R}(\Lambda_c)$ (identified with the quotient $\Lambda_f/\Lambda_c$) and $\Y$ the corresponding output distribution. Consider the function $\psi(\rho)=\psi(\rho|W,\mathcal{U}_{\mathcal{X}})$ in Definition \ref{Hayashi_function}. 
		From (\ref{eq_psi_0}) and (\ref{eq_psi_2}), it follows that its Taylor expansion in $0$ is given by
		\begin{equation} \label{taylor}
		\psi(\rho)=\rho \mathbb{I}(\X;\Y)+\rho^2 \psi''(0)+o(\rho^2),
		\end{equation}
		where $\psi''(0)$ is given in Lemma \ref{second_derivative}. Noting that
		{\allowdisplaybreaks
		\begin{align*}
		&(W \circ \mathcal{U}_{\mathcal{X}})(y)=\sum_{x \in \mathcal{X}} \frac{1}{\abs{\mathcal{X}}} W_x(y)\\
		&=
		\sum_{x \in \Lambda_f/\Lambda_c} \frac{1}{\abs{\Lambda_f/\Lambda_c}}f_{\sigma,\Lambda_c}(y-x)
		=\frac{1}{\abs{\Lambda_f/\Lambda_c}}f_{\sigma,\Lambda_f}(y),
		\end{align*}
		}
		we find that $\psi''(0)$ is equal to
		{\allowdisplaybreaks
		\begin{align*}
		&\sum_{x \in \Lambda_f/\Lambda_c}\!\!\frac{1}{\abs{\Lambda_f/\Lambda_c}}\! \int_{\mathcal{R}(\Lambda_c)} \!f_{\sigma,\Lambda_c}(y\!-\!x)\bigg[\!\log \smash{\frac{f_{\sigma,\Lambda_c}(y\!-\!x)}{\frac{1}{\abs{\Lambda_f/\Lambda_c}}f_{\sigma,\Lambda_f}(y)}}\!\bigg]^2\!\!dy\\
		&\!-\!
		\!\Bigg[\sum_{x \in \Lambda_f/\Lambda_c}\!\!\frac{1}{\abs{\Lambda_f/\Lambda_c}}\!\int_{\mathcal{R}(\Lambda_c)}\!\! \!f_{\sigma,\Lambda_c}\!(y\!-\!x)\log\! \frac{f_{\sigma,\Lambda_c}\!(y\!-\!x)}{\frac{1}{\abs{\Lambda_f/\Lambda_c}}f_{\sigma,\Lambda_f}\!(y)}dy\Bigg]^2\\
		&\leq\! \!\! \!\sum_{x \in \Lambda_f/\Lambda_c}\!\!\frac{1}{\abs{\Lambda_f/\Lambda_c}}\!\int_{\mathcal{R}(\Lambda_c)}\! f_{\sigma,\Lambda_c}\!(y\!-\!x)\!\left[\log \frac{f_{\sigma,\Lambda_c}(y\!-\!x)}{\frac{1}{\abs{\Lambda_f/\Lambda_c}}f_{\sigma,\Lambda_f}\!(y)}\right]^2\!\!\!dy\\
		&=\int_{\mathcal{R}(\Lambda_c)}\! f_{\sigma,\Lambda_c}(y')\left(\log \frac{f_{\sigma,\Lambda_c}(y')}{\frac{1}{\abs{\Lambda_f/\Lambda_c}}f_{\sigma,\Lambda_f}(y')}\right)^2dy'
		\end{align*}
		}%
		with the change of variables $y'=y-x \Mod \mathcal{R}(\Lambda_c)$. From the definition of flatness factor and the bound (\ref{flatness_factor_bound}), we find that $\forall y'\in \mathcal{R}(\Lambda_c)$,
		$$f_{\sigma,\Lambda_f}(y')\geq \frac{1-\epsilon_{\Lambda_f}(\sigma)}{V(\Lambda_f)}\geq \frac{1-4 e^{-\frac{2 \pi^2 \sigma^2}{\alpha^2}}}{\alpha}.$$
		Recalling the definition of the theta series of a lattice in  (\ref{theta_series}) and the relation (\ref{theta_dual}), we have 
		$\epsilon_{\Lambda}(\sigma)=\Theta_{\Lambda^*}(2\pi \sigma^2)-1$, where $\Lambda^*$ is the dual lattice of $\Lambda$. Then by \cite[Remark 1]{LLBS}, $\forall y'\in \mathcal{V}(\Lambda_c)$
		\begin{align*}
		&f_{\sigma,\Lambda_c}(y')\leq f_{\sigma,\Lambda_c}(0)=\frac{1}{\sqrt{2\pi}\sigma}\Theta_{\Lambda_c}\left(\frac{1}{2\pi\sigma^2}\right)\\
		&=\frac{1}{\sqrt{2\pi}\sigma}\left(1+\epsilon_{\Lambda_c^*}\left(\frac{1}{2 \pi \sigma}\right)\right).
		\end{align*}
		Again using the bound (\ref{flatness_factor_bound}), we have
		$$\epsilon_{\Lambda_c^*}\left(\frac{1}{2 \pi \sigma}\right)=\epsilon_{\frac{1}{\alpha p}\mathbb{Z}}\left(\frac{1}{2 \pi \sigma}\right)\leq 4 e^{-\frac{\alpha^2p^2}{2\sigma^2}}.$$
		Then, since $\alpha \to 0$ and $\alpha p \to \infty$ when $n \to \infty$, for sufficiently large $n$ we have 
		$$\frac{f_{\sigma,\Lambda_c}(y')}{\frac{1}{\abs{\Lambda_f/\Lambda_c}}f_{\sigma,\Lambda_f}(y')} \leq \frac{1}{\sqrt{2\pi}\sigma} \frac{\alpha p (1+4 e^{-\frac{\alpha^2 p^2}{2\sigma^2}})}{1-4e^{-\frac{2 \pi^2\sigma^2}{\alpha^2}}} \leq C \alpha p$$
		for some constant $C>0$. Consequently, for large enough $n$, $\exists C'>0$ such that $$\psi''(0) \leq C' (\log \alpha p)^2.$$
	Then, from the Taylor expansion (\ref{taylor}) we obtain the bound
		$$\psi(\rho) \leq \rho \mathbb{I}(\X;\Y)+\rho^2 C'' (\log \alpha p)^2$$
		for another suitable constant $C''>0$. 
		In particular, we can bound the exponent in equation (\ref{E_Fn}) as follows:
		$$\rho R -\psi(\rho|W, \mathcal{U}_{\mathcal{X}}) \geq \rho (R-\mathbb{I}(\X;\Y)-\rho C''(\log \alpha p)^2)>\rho \frac{\delta_0}{2}$$
		for sufficiently large $n$, where $\delta_0$ is defined in (\ref{delta_0}), as long as $\rho=o\left(\frac{1}{(\log \alpha p)^2}\right)$ and the VNR condition (\ref{VNR_condition}) is satisfied.
		In particular if we choose the scaling\footnote{This choice of scaling is compatible with the existence of a suitable sequence of nested lattices, see Appendix \ref{existence_section}.} 
		\begin{equation} \label{scaling}
		p=\xi n^{3/2}, \quad \alpha p=2 \sqrt{n},
		\end{equation}
		where $\xi$ is the smallest number in the interval $[1,2)$ such that $p$ is prime \cite[Section IV]{Ordentlich_Erez_16}, we have convergence in (\ref{rho_bar}) with $\bar{\rho}=\frac{1}{(\log 2\sqrt{n})^{2+\eta}}$ for some $\eta>0$ since
		\begin{align*}
		\frac{1}{\bar{\rho}}e^{-n\bar{\rho}\frac{\delta_0}{2}}=(\log 2\sqrt{n})^{2+\eta} e^{-\frac{n\delta_0}{2(\log 2\sqrt{n})^{2+\eta}}} \to 0. 
		\end{align*}
This concludes the proof of Remark \ref{remark_convergence}.
\end{IEEEproof}
			
			
 Then according to Corollary \ref{corol_resolvability}, 
 for $\mathsf{C}_n$ chosen uniformly in the set of
 $(n,k)$ linear codes 
 over $\mathbb{F}_p$ of rate $R=\frac{k}{n}\log p$,  
			$$\mathbb{E}_{\mathsf{C}_n}\left[\mathbb{D}(W^n \circ \mathcal{U}_{\mathsf{C}_n}\Vert W^n \circ \mathcal{U}_{\mathcal{X}}^{\otimes n})\right] \leq \frac{1}{\bar{\rho}}e^{-n\bar{\rho}\frac{\delta_0}{2}} \to 0$$ as $n \to \infty$. In particular, there exists at least one code $\mathcal{C}_n$ such that $\mathbb{D}(W^n \circ \mathcal{U}_{\mathcal{C}_n}\Vert W^n \circ \mathcal{U}_{\mathcal{X}}^{\otimes n})\to 0.$
			Note that
			{\allowdisplaybreaks
			\begin{align*}
			& (W^n \circ \mathcal{U}_{\mathcal{C}_n})(\mb{y})=\sum_{\mb{c}\in\mathcal{C}_n} \frac{1}{\abs{\mathcal{C}_n}}f_{\sigma,\Lambda_c^n}(\mb{y}-\alpha \mb{c})\\
			&=\sum_{\mb{c}\in\mathcal{C}_n} \sum_{\boldsymbol\lambda_c \in \Lambda_c^n} \frac{1}{p^k} f_{\sigma}(\mb{y}-\alpha \mb{c} -\boldsymbol\lambda_c)=\frac{1}{p^k}\sum_{\boldsymbol\lambda \in \Lambda} f_{\sigma}(\mb{y}-\boldsymbol\lambda) \\
			&=\frac{1}{p^k} f_{\sigma,\Lambda}(\mb{y}). 
			\end{align*}
			}
			On the other hand,
			{\allowdisplaybreaks
			\begin{align*}
			& (W^n \circ \mathcal{U}_{\mathcal{X}}^{\otimes n})(\mb{y})=
				\sum_{\mb{x} \in \Lambda_f^n \cap \mathcal{R}(\Lambda_c^n)} \frac{1}{p^n} f_{\sigma, \Lambda_c^n}( \mb{y}-\mb{x})\\
				&=\frac{1}{p^n} f_{\sigma, \Lambda_f^n}(\mb{y}) 
				.
			\end{align*}
			}
						Since both $(W^n \circ \mathcal{U}_{\mathcal{C}_n})$ and $(W^n \circ \mathcal{U}_{\mathcal{X}}^{\otimes n})$ are $\Lambda$-periodic, we can write
						{\allowdisplaybreaks
			\begin{align*}
			&\mathbb{D}(W^n \circ \mathcal{U}_{\mathcal{C}_n}\Vert W^n \circ \mathcal{U}_{\mathcal{X}}^{\otimes n})\\
			&=\int_{\mathcal{R}(\Lambda_c^n)} p^{-k} f_{\sigma,\Lambda}(\mb{y}) \log \frac{p^{-k} f_{\sigma,\Lambda}(\mb{y})}{p^{-n}f_{\sigma,\Lambda_f^n}(\mb{y})} d\mb{y}\\
			&= \int_{\mathcal{R}(\Lambda)} f_{\sigma,\Lambda}(\mb{y}) \log \frac{ f_{\sigma,\Lambda}(\mb{y})}{p^{-(n-k)}f_{\sigma,\Lambda_f^n}(\mb{y})} d\mb{y}\\
			&=
			\mathbb{D}(f_{\sigma,\mathcal{R}(\Lambda)}\Vert p^{-(n-k)} {f_{\sigma,\Lambda_f^n}}_{|\mathcal{R}(\Lambda)})=C(\Lambda_f^n/\Lambda,\sigma^2) \to 0
			\end{align*}
			}
			using Lemma \ref{capacity_lemma}. This concludes the proof. 
\end{IEEEproof}
		
		\begin{rem} \label{measure_remark}
		With a standard argument based on Markov's inequality, we can also show that the set of KL-secrecy good lattices has large measure, since $\forall \xi>0$, 
		\begin{align*}
		&\mathbb{P}\left\{\mathbb{D}(W^n \circ \mathcal{U}_{\mathsf{C}_n}\Vert W^n \circ \mathcal{U}_{\mathcal{X}}^{\otimes n})>\xi\right\}\\
		&\leq \frac{1}{\xi} \mathbb{E}_{\mathsf{C}_n}\left[\mathbb{D}(W^n \circ \mathcal{U}_{\mathsf{C}_n}\Vert W^n \circ \mathcal{U}_{\mathcal{X}}^{\otimes n})\right].
		\end{align*}
		Given $0<c<1/2$, we can take $\xi=\frac{1}{c}\frac{e^{-n\bar{\rho}\delta_0}}{\bar{\rho}}$ and we obtain 
		$$\mathbb{P}\left\{\mathbb{D}(W^n \circ \mathcal{U}_{\mathsf{C}_n}\Vert W^n \circ \mathcal{U}_{\mathcal{X}}^{\otimes n})>\xi\right\}\leq c.$$
		\end{rem}

\section{Existence of a sequence of nested lattices for secret key generation} \label{existence_section}
In this section, we show the existence of a sequence of nested lattices $\Lambda_3^{(n)} \subset \Lambda_2^{(n)} \subset \Lambda_1^{(n)}$ such that $\Lambda_3$ is KL secrecy-good, $\Lambda_2$ is AWGN-good and $\Lambda_1$ is KL secrecy-good. By Remark \ref{Pinsker2}, it follows that
$\Lambda_1$ and $\Lambda_3$ are also $L^1$-secrecy good. Note that we don't need covering-goodness, which requires more stringent conditions on the parameters \cite{Erez_Litsyn_Zamir}. 

We will follow the construction in \cite{Ordentlich_Erez_16}. We denote by  $V_{\mathcal{B},n}$ the volume of the $n$-dimensional ball of radius $1$. Given $P_3>P_2>P_1>0$, let $a_i=\log \frac{1}{P_i}$ for $i=1,2,3$. We consider the dimensions $k_3 < k_2 < k_1 \leq n$ defined as follows:
$$k_i=\floor{\frac{n}{2\log p} \left(\log\left(\frac{4}{V_{\mathcal{B},n}^{2/n}}\right)+a_i\right)}, \quad i=1,2,3,$$
where $p=\xi n^{3/2}$, and $\xi$ is taken to be the smallest number in the interval $[1,2)$ such that $p$ is prime \cite[Section IV]{Ordentlich_Erez_16}\footnote{Note that the conclusions of \cite{Ordentlich_Erez_16} still hold for any $p=\Theta(n^{\frac{1}{2}+\delta})$ with $\delta>0$, see Remark 7 in that paper.}.
Let $\mathcal{C}_1$ be uniformly sampled from the set of all linear $(n,k_1)$ codes over $\mathbb{F}_p$, with generator matrix $G_1$ (in column notation). We denote by $G_2$ and $G_3$ the submatrices of $G_1$ corresponding to the first $k_2$ and $k_3$ columns respectively, and by $\mathcal{C}_2$, $\mathcal{C}_3$ the corresponding linear codes.  
Finally, we define the lattices $\tilde{\Lambda}_i=\frac{1}{p}\mathcal{C}_i+\mathbb{Z}^n$ and $\Lambda_i=\alpha p \tilde{\Lambda}_i$ for $i=1,2,3$, where $\alpha=\frac{2\sqrt{n}}{p}$.
Then by \cite[Theorem 1 and Theorem 6]{Ordentlich_Erez_16}, the matrices $G_1$, $G_2$, $G_3$ are full rank and  the nested lattices $\Lambda_3^{(n)} \subset \Lambda_2^{(n)} \subset \Lambda_1^{(n)}$ obtained in this way are good for quantization and coding with probability that tends to $1$ when $n \to \infty$ and 
$$\lim_{n \to \infty} V^{2/n}(\Lambda_i^{(n)})=2 \pi e P_i, \quad i=1,2,3.$$
 Note that we have taken the same scaling as in (\ref{scaling}). In particular, when $n \to \infty$ we have $p \to \infty$, $\alpha \to 0$ and $\alpha p \to \infty$.
 
  Moreover, $\alpha=\frac{2}{\xi n}$ satisfies the condition $\alpha =o\left(\frac{1}{n^{c}}\right)$
in Appendix \ref{proof_VNR_L1}.
Therefore, due to Remark \ref{measure_remark} the lattices $\Lambda_3$ and $\Lambda_1$ are also KL secrecy-good with probability close to $1$, which concludes the proof.

\section{Optimal public rate / secret key rate trade-off} \label{Appendix_tradeoff}
In this section, we derive the optimal trade-off between public rate and secret key rate from \cite{WaOh10} for the setting in our paper. Note that Theorem 4 in \cite{WaOh10} doesn't directly apply to our model because our source doesn't necessarily satisfy $\X \to \Y \to \Z$. However, the proof of Lemma 6 in \cite{WaOh10} shows how to obtain a new source $(\bar{\X}, \bar{\Y}, \bar{\Z})$ which is degraded ($\bar{\X} \to \bar{\Y} \to \bar{\Z}$) and has the same achievable region ($\mathcal{R}(\X,\Y,\Z)=\mathcal{R}(\bar{\X}, \bar{\Y}, \bar{\Z})$). In particular, translating the proof into our notation, we can take $\bar{\X}=\X$, $\bar{\Y}=\Y$ and 
$$\bar{\Z}=\frac{\sigma_z \rho_{xz}}{\sigma_y \rho_{xy}} \Y + \hat{\mathsf{N}},$$
where $\hat{\mathsf{N}}$ is independent of all other random variables and has variance $\sigma_z^2\left(1 - \frac{\rho_{xz}^2}{\rho_{xy}^2}\right)$. \\
From elementary computations we see that $\sigma_{\bar{z}}=\sigma_z$, $\rho_{x\bar{z}}=\rho_{xz}$ and $\rho_{y\bar{z}}=\frac{\rho_{xz}}{\rho_{xy}}$. \\
In our notation, the optimal trade-off given by Theorem 4 of \cite{WaOh10} is given by
$$ R_K \leq \frac{1}{2} \log \frac{(1-\rho_{\bar{y}\bar{z}}^2)(1-\rho_{\bar{x}\bar{z}}^2)-(\rho_{\bar{x}\bar{y}}-\rho_{\bar{y}\bar{z}}\rho_{\bar{x}\bar{z}})^2 e^{-2R_P}}{(1-\rho_{\bar{y}\bar{z}}^2)(1-\rho_{\bar{x}\bar{z}}^2)-(\rho_{\bar{x}\bar{y}}-\rho_{\bar{y}\bar{z}}\rho_{\bar{x}\bar{z}})^2}.$$
In terms of the original variables $\X, \Y, \Z$, after simplifying the expression we obtain the optimal trade-off
\begin{equation*}
R_K \leq \frac{1}{2} \log \frac{(1-\rho_{xz}^2)-(\rho_{xy}^2-\rho_{xz}^2)e^{-2R_P}}{1-\rho_{xy}^2}. 
\end{equation*}
(Recall that $\rho_{xy} > \rho_{xz}$ in our setting). Using the notation $\sigma_1^2=\sigma_x^2(1-\rho_{xy}^2)$, $\sigma_2^2=\sigma_x^2(1-\rho_{xz}^2)$ from our paper, this is equal to
\begin{equation}
R_K \leq \frac{1}{2} \log \left( e^{-2R_P} + \frac{\sigma_2^2}{\sigma_1^2} (1 - e^{-2R_P})\right). 
\end{equation}

\section{Proof of Lemma \ref{second_derivative}} \label{proof_second_derivative}
The first derivative of the function $\psi(\rho)=\psi(\rho|W,p_{\X})$ is
{\allowdisplaybreaks
\begin{align*}
&\psi'(\rho)=
\frac{\sum_{x \in \mathcal{X}} p_{\X}(x) \int_{\mathcal{Y}} \frac{{W_x(y)}^{1+\rho}}{\left((W \circ p_{\X})(y)\right)^{\rho}} \log\frac{W_x(y)}{(W \circ p_{\X})(y)}dy}{\sum_{x \in \mathcal{X}} p_{\X}(x) \int_{\mathcal{Y}} \frac{W_x(y)^{1+\rho}}{\left((W \circ p_{\X})(y)\right)^{\rho}} dy}\\
&=\frac{f(\rho)}{g(\rho)}.
\end{align*}
}
Then we have
\begin{align*}
&g(0)=1,\\
&f(0)=\sum_{x \in \mathcal{X}} p_{\X}(x) \int_{\mathcal{Y}} W_x(y) \log\frac{W_x(y)}{(W \circ p_{\X})(y)}dy=g'(0),\\
&f'(0)=\sum_{x \in \mathcal{X}} p_{\X}(x) \int_{\mathcal{Y}} W_x(y) \left(\log\frac{W_x(y)}{(W \circ p_{\X})(y)}\right)^2dy.
\end{align*}
The conclusion follows since
\begin{align}
&\psi''(0)=\frac{f'(0)g(0)-f(0)g'(0)}{g^2(0)}. \tag*{\IEEEQED}
\end{align}

\small
\bibliographystyle{IEEEtran}
\bibliography{IEEEabrv,wiretap}

\begin{IEEEbiographynophoto}{Laura Luzzi} received the degree in 
Mathematics from the University of Pisa, Italy, in 2003 and the Ph.D. 
degree in Mathematics for Technology and Industrial Applications from 
Scuola Normale Superiore, Pisa, Italy, in 2007. From 2007 to 2012 she held 
postdoctoral positions in T\'el\'ecom-ParisTech and Sup\'elec, France, and 
a Marie Curie IEF Fellowship at Imperial College London, United Kingdom. 
Since 2012, she is an Assistant Professor at ENSEA, Cergy-Pontoise, 
France, and a researcher at ETIS (UMR 8051, CY Cergy Paris Université, ENSEA, CNRS).\\ 
Her research interests include coding for wireless 
communications, physical layer security and lattice-based cryptography. 
\end{IEEEbiographynophoto} 

\begin{IEEEbiographynophoto}{Cong Ling}
received the B.S. and M.S. degrees in electrical engineering from
the Nanjing Institute of Communications Engineering, Nanjing, China,
in 1995 and 1997, respectively, and the Ph.D. degree in electrical
engineering from the Nanyang Technological University, Singapore, in
2005.
He is currently a Reader in the Electrical and Electronic
Engineering Department at Imperial College London. His research
interests are information theory and applied mathematics, with a focus on lattices.
Dr. Ling has served as an Associate Editor of IEEE Transactions on Communications and of IEEE Transactions on Vehicular Technology.
\end{IEEEbiographynophoto}

\begin{IEEEbiographynophoto}{Matthieu R. Bloch} is a Professor in the School of Electrical and Computer Engineering. He received the Engineering degree from Supélec, Gif-sur-Yvette, France, the M.S. degree in Electrical Engineering from the Georgia Institute of Technology, Atlanta, in 2003, the Ph.D. degree in Engineering Science from the Université de Franche-Comté, Besançon, France, in 2006, and the Ph.D. degree in Electrical Engineering from the Georgia Institute of Technology in 2008. In 2008-2009, he was a postdoctoral research associate at the University of Notre Dame, South Bend, IN. Since July 2009, Dr. Bloch has been on the faculty of the School of Electrical and Computer Engineering, and from 2009 to 2013 Dr. Bloch was based at Georgia Tech Europe. His research interests are in the areas of information theory, error-control coding, wireless communications, and cryptography. Dr. Bloch has served on the organizing committee of several international conferences; he was the chair of the Online Committee of the IEEE Information Theory Society from 2011 to 2014, an Associate Editor for the IEEE Transactions on Information Theory from 2016 to 2019 and again since 2021, and he has been on the Board of Governors of the IEEE Information Theory Society since 2016 and currently serves as the President. He was an Associate Editor for the IEEE Transactions on Information Forensics and Security from 2019 to 2023. He is the co-recipient of the IEEE Communications Society and IEEE Information Theory Society 2011 Joint Paper Award and the co-author of the textbook Physical-Layer Security: From Information Theory to Security Engineering published by Cambridge University Press.
\end{IEEEbiographynophoto}

%





\end{document}